\documentclass[aps,prd,showpacs,eqsecnum,twocolumn,superscriptaddress,nofootinbib]{revtex4}

\usepackage{latexsym} \usepackage{amssymb} \usepackage{amsfonts}
\usepackage{amsmath} \usepackage{bm} \usepackage[dvips]{graphicx}
\usepackage{color}
\usepackage{subfigure} \usepackage{times} \usepackage{units}
\usepackage{hyperref}
\usepackage{bm}
\usepackage[utf8x]{inputenc} \usepackage{amssymb,amsmath}
\usepackage{graphicx}

\begin{document}
\title{Application of dictionary learning to denoise LIGO's blip noise transients}

\author{Alejandro \surname{Torres-Forn\'e}}
\affiliation{Max Planck Institute for Gravitational Physics (Albert Einstein Institute), D-14476 Potsdam-Golm, Germany}
 
\author{Elena \surname{Cuoco}}
\affiliation{European Gravitational Observatory (EGO), Via E Amaldi, I-56021 Cascina, Italy}
\affiliation{Scuola Normale Superiore (SNS), Piazza dei Cavalieri, 7, 56126 Pisa PI, Italy}
\affiliation{ Istituto Nazionale di Fisica Nucleare (INFN) Sez. Pisa Edificio C - Largo B. Pontecorvo 3, 56127 Pisa, Italy.}

\author{Jos\'e A. \surname{Font}}\affiliation{Departamento de
 Astronom\'{\i}a y Astrof\'{\i}sica, Universitat de Val\`encia,
 Dr. Moliner 50, 46100, Burjassot (Val\`encia), Spain}
 \affiliation{Observatori Astron\`omic, Universitat de Val\`encia, C/ Catedr\'atico 
 Jos\'e Beltr\'an 2, 46980, Paterna (Val\`encia), Spain} 
 
 \author{Antonio \surname{Marquina}}\affiliation{Departamento de
 Matem\'atica Aplicada, Universitat de Val\`encia,
 Dr. Moliner 50, 46100, Burjassot (Val\`encia), Spain} 
 

\begin{abstract}
Data streams of gravitational-wave detectors are polluted by transient noise features, or ``glitches'', of instrumental and environmental origin. In this work we investigate the use of total-variation methods and learned dictionaries to mitigate the effect of those transients in the data. We focus on a specific type of transient, ``blip" glitches, as this is the most common type of glitch present in the LIGO detectors and their waveforms are easy to identify. We randomly select 100 blip glitches scattered in the data from advanced LIGO's O1 run, as provided by the citizen-science project Gravity Spy. Our results show that dictionary-learning methods are a valid approach to model and subtract most of the glitch contribution in all cases analyzed, particularly at frequencies below $\sim 1$ kHz. The high-frequency component of the glitch is best removed when a combination of dictionaries with different atom length is employed. As a further example we apply our approach to the glitch visible in the LIGO-Livingston data around the time of merger of binary neutron star signal GW170817, finding satisfactory results. This paper is the first step in our ongoing program to automatically classify and subtract all families of gravitational-wave glitches employing variational methods. 
\end{abstract}
\maketitle

\section{Introduction}
\label{section:intro}

%
%
The third observational campaign of the advanced gravitational-wave (GW) 
detectors LIGO~\cite{AdvLIGO} and Virgo~\cite{Virgo}, O3, 
is currently ongoing. During the previous two campaigns, O1 and O2, the 
GW detector network reported the observation of eleven compact binary 
mergers~\cite{GWTC-1} comprising ten binary black holes and one binary 
neutron star. The latter, GW170817~\cite{GW170817}, was 
accompanied by extensive and very successful follow-up observations of 
electromagnetic emission originated from the same astronomical 
source~\cite{MMA}. Moreover, neutrino searches were also carried 
out, yet no detection has been reported~\cite{neutrino}. The entire GW 
strain data from O1/O2 has been made publicly available and, in 
particular, the data around the time of each of the eleven O1/O2 events 
are accessible through the Gravitational-Wave Open Science 
Center~\footnote{{\url{https://gw-openscience.org}}}. Since the start of O3 on April 1st 2019, GW candidate 
events are being released as public alerts to facilitate the rapid identification 
of electromagnetic or neutrino counterparts. The growing list of candidates 
can be inspected at the GW Candidate Event Database~\cite{GraceDB}. 
Toward the end of O3, the GW detector network may be increased by yet 
another facility with the addition of the KAGRA detector~\cite{KAGRA}.

%
%

The detection of GWs is severely hampered by many sources of noise that 
contribute to a non-stationary background in the time series of data in 
which actual GW signals reside. The sensitivity of the instruments is 
limited at low frequencies (below $\sim 20$ Hz) by gravity-gradient 
(seismic) noise and at high frequencies (above $\sim$ 2 kHz) by 
photon-shot noise originated by quantum fluctuations of the laser. 
The detectors are most sensitive at intermediate frequencies ($\sim 200$ 
Hz) where the Brownian motion of the suspensions and mirrors is the 
limiting source of so-called thermal noise. Moreover, the data stream is 
polluted with the presence of transient (short duration) noise 
signals, commonly known as ``glitches'', whose origin is not 
astrophysical but rather instrumental and environmental. We refer 
to~\cite{LIGO-Virgo-guide} for a comprehensive overview of the 
LIGO/Virgo detector noise and the extraction of GW signals.

Glitches difficult GW data analysis for a number of reasons. By their 
short-duration nature they contribute significantly to the background of 
transient GW searches. Glitches may occur sufficiently frequently to potentially affect true 
signals, particularly when occurring in (or almost in) coincidence. Furthermore, some 
types of glitches show time-frequency morphologies remarkably similar to 
actual transient astrophysical signals, which increases the false-alarm 
rate of potential triggers. Moreover, having to remove portions of data 
in which glitches are present downgrades the duty cycle of the 
detectors. It is however not trivial to remove defective segments of data. 
The simplest approach, i.e.~setting them to zero, might result in a leakage of 
excess power, which may turn the mitigating approach more damaging than the 
very effect of the glitch. 

For all these reasons, understanding the origin of glitches 
and mitigating their effects is a major effort in the characterization of GW 
detectors~\cite{glitch-LVC-1,glitch-LVC-2}. Indeed, in recent years many strategies 
have been developed to automatically classify glitches. The approaches are as 
diverse as Bayesian inference, machine learning, deep learning, and citizen 
science~\cite{Powell:2015, Powell:2017, Zevin:2017, Mukund:2017, George:2018, 
Razzano:2018, Miquel:2019, Coughlin:2019, Colgan:2019}. 
Recent examples of glitch mitigation are reported in~\cite{Pankow:2018,Zackay:2019a,Venumadhav:2019,Wei:2020}. Ref.~\cite{Pankow:2018} describes various deglitching methods to extract the strong glitch present in the LIGO-Livingston detector about 1s before the merger of the binary neutron star that produced the signal GW170817~\cite{GW170817}. In~\cite{Zackay:2019a,Venumadhav:2019} the 
impact of loud glitches is reduced by using an inpainting filter that fills the hole created after 
windowing the glitch. Glitch reduction, together with other techniques, was shown 
to improve the statistical significance of a GW trigger. In addition, deep learning approaches have also proven very effective to recover the true GW signal even in 
the presence of glitches \cite{Wei:2020}.

There are many different families of glitches identified during the 
advanced LIGO-Virgo observing runs \cite{Zevin:2017, Cabero:2019, Nitz:2018}. 
Glitches from each family have a similar morphology, 
although the characteristics of each specific glitch in terms of 
duration, bandwidth and signal-to-noise (SNR) ratio can vary 
significantly even for glitches inside the same family. In this work we 
focus on blip glitches, a noise transient characterised by a duration of about 10 ms and a 
frequency bandwidth of about 100 Hz. This type of glitches, which has 
mainly been found in the two LIGO detectors, significantly reduce the 
sensitivity of searches for high-mass compact binary coalescences. 
Blip glitches in LIGO data are identified using both the PyCBC pipeline 
search (see \cite{Cabero:2019} and references therein) and the citizen-science 
effort Gravity Spy~\cite{Zevin:2017}. The recent study of 
\cite{Cabero:2019} based on PyCBC has found that Advanced LIGO data during 
O1/O2 contains approximately two blip glitches per hour of data 
(amounting to thousands of blip glitches in total). The physical 
origin of most of them remains unclear.

%
%
This paper explores the performance of dictionary-learning methods~\cite{Mairal:2009} 
to mitigate the presence of blip glitches in advanced LIGO data. To this 
aim we select a large number of blip glitches randomly distributed along 
the data stream from advanced LIGO’s first observing run. For each 
glitch, the data correspond to a one-second window centered at the GPS 
time of the glitch as provided by Gravity Spy~\cite{Zevin:2017}. As in~\cite{Pankow:2018,Zackay:2019a,Wei:2020}
the goal of our work is to mitigate the impact of glitches in GW data in order to increase the 
statistical significance of astrophysical triggers. Our results show that dictionary-learning
techniques are able to model blip glitches and to subtract them from the data 
without significantly disturbing the background.

%
%
The paper is organized as follows: In Section~\ref{section:methods} we summarize the mathematical 
framework of the variational methods which are at the core of the dictionary-learning approach we use. 
Section~\ref{section:setup} discusses technical aspects, namely the whitening procedure we employ 
to remove noise lines and other artefacts, the training of the dictionaries, and how we perform the 
reconstruction of the glitches. The results of our study are presented in Section~\ref{section:results}. 
Finally, a summary is provided in Section~\ref{section:summary}. Appendix A shows the spectrograms 
of the 16 blip glitches from O1 we employ in our test set and reports their main characteristics.

\section{Review of $L_1$-norm variational methods} 
\label{section:methods}

\subsection{Total variation methods}

In this paper we employ two different variational techniques based on the $\text{L}_1$ norm,
the Rudin-Osher-Fatemi (ROF) method~\cite{Rudin:1992} and a Dictionary Learning 
method~\cite{Mairal:2009}. We have recently begun to use these procedures in the context 
of GW data analysis in~\cite{Torres:2014, Torres:2016, Torres:2018, Miquel:2019}. Both approaches 
solve the denoising problem, $ y=u+n$, where $u$ is the true signal and $n$ is the noise, as a variational 
problem. The solution $u$ is thus obtained as
\begin{equation}
 \label{eq:general}
 u_{\lambda}=\underset{u} {\text{argmin}}\left\{{\cal R}(u)+\frac{\lambda}{2} {\cal F} (u) \right\}~,
 \end{equation}
where ${\cal R}$ is the regularization term, i.e.~the constrain to impose in the data and ${\cal F}$ is the fidelity term, which measures the similarity of the solution to the data. The parameter $\lambda$ is the regularization parameter and controls the relative weight of both terms in the equation. Even though both methods solve the same general problem, each one of them approaches the problem in a different way and, therefore, the regularization term and the fidelity term have different expressions.
 
In 1992, Rudin, Osher and Fatemi \cite{Rudin:1992} proposed the use of the so.called total-variation (TV) norm as 
the regularization term ${\cal R}(u) = \int_\Omega |\nabla u|$ constrained to $||y-u||^2$. Note that $|\cdot |$ and $||\cdot||$ represent the $\text{L}_1$ and $\text{L}_2$ norms, respectively. This specific formulation of the variational problem~(\ref{eq:general}) is called ROF model and reads
\begin{equation}
 \label{eq:TV_model}
u_{\lambda}=\underset{u} {\text{argmin}}\left\{ \int_\Omega |\nabla u|+\frac{\lambda}{2} \, ||y-u||^2\right\}~.
\end{equation}
This model preserves steep gradients, reduces noise by {\it sparsifying} (i.e.~promoting zeros) the gradient of the signal and 
avoids spurious oscillations (Gibbs effect). However, the associated Euler-Lagrange equation, given by
\begin{equation}
\label{eq:EL_TV}
\nabla \cdot \frac{\nabla u }{|\nabla u|}+ \lambda (y-u) = 0~,
\end{equation}
becomes singular when $|\nabla u| = 0$. This issue can be easily solved by changing the standard 
TV norm by a slightly perturbed version (see ~\cite{Torres:2014} for a detailed explanation),
\begin{equation}
\mathrm{TV}_{\beta}(u):= \int \sqrt{|\nabla u|+\beta}~,
\end{equation}
where $\beta$ is a small positive parameter. We refer to this modified version of the regularization term in the method as regularized ROF (rROF).

\subsection{Sparse reconstruction over a fixed dictionary}

In dictionary-based methods, the denoising is performed by assuming that the true signal $u$ can be represented as a linear
combination of the columns (atoms) of a matrix $\bm{D}$ called the dictionary. If the signal 
can be represented with a {\it few} columns of $\bm{D}$, the dictionary is adapted to $u$. In other words, there exists a ``sparse vector'' $\alpha$ such that $u \sim \bm{D} \alpha$. As a result, the fidelity term in Eq.~(\ref{eq:general}) reads,
\begin{equation}
{\cal F}(\alpha) = ||y-\bm{D}\alpha||^2\,.
\end{equation}

In other words, the problem reduces to finding a sparse vector $\alpha$ that represents the signal $u$ over the columns of the dictionary. 
The next step is to find a regularisation term that induces sparsity over the coefficients of $\alpha$. 
Classical dictionary-learning techniques \citep{Olshausen:1997, Aharon:2006} use as regularisation term the $\rm{L}_0$-norm, which is chosen to ensure that the solution has the fewest possible number of nonzero coefficients. However, this problem is not convex and is NP-hard, i.e.~it can be solved in non-deterministic polynomial-time.
If the $\rm{L}_1$-norm is used instead of the $\rm{L}_0$-norm, the problem becomes convex. This type of regularisation promotes zeros in the components of the vector coefficient $\alpha$, and the solution is the sparsest one in most cases. 
The variational problem thus reads,
\begin{equation}
\label{eq:lasso}
\alpha_{\lambda}=\underset{\alpha}{\rm{argmin}} \left\{ |\alpha|+\frac{\lambda}{2}||\bm{D}\alpha- y||^2\right\},
\end{equation}
which is known as {\it basis pursuit}~\citep{Chen:2001} or {\it LASSO}~\citep{lasso}. In this paper we solve Eq.~(\ref{eq:lasso}) using the Alternating Direction Method of Multipliers (ADMM) algorithm~\cite{Boyd:2011}.

\subsection{Dictionary Learning}

In the previous section we have assumed that the dictionary $\bm{D}$ is fixed and we only solve the problem of representation. Traditionally, predefined dictionaries based on wavelets, curvelets, etc, have been used. However, signal reconstruction can be dramatically improved by learning the dictionary instead of using a predefined one~\cite{Elad:2006}.
In this approach a set of training signals is divided into patches in such a way that the length of the patches is less than the total length of the training signals. In most common problems, the number of training patches $m$ is large compared with the length of each patch $n$, $n\ll m$. The procedure to train the dictionary is similar to Eq.~(\ref{eq:lasso}) except that the dictionary $\bm{D}$ should now be added as variable,
\begin{equation}
\label{eq:dict_learning}
\alpha_{\lambda}, \bm{D}_{\lambda}=\underset{\alpha, \bm{D}}{\rm{argmin}} \left\{\frac{1}{n}\sum_{i=1}^{m}||\bm{D}\alpha_i- {x}_i||^2_2+\lambda|\alpha_i|\right\},
\end{equation}
where $x_i$ denotes the $i$-th training patch. Unfortunately, this problem is not jointly convex unless the variables are considered separately. In~\cite{Mairal:2009} a method based on stochastic approximations was proposed. These approximations process one sample at a time and the method takes advantage of the problem structure to efficiently solve it. 
For each element in the training set, the algorithm alternates a classical sparse coding step, to solve for $\alpha$ using a dictionary $\bm{D}$ obtained in the previous iteration, with a dictionary update step, where the new dictionary is calculated with the recently calculated values of $\alpha$, namely
\begin{eqnarray}
\alpha^{k+1}&=&\underset{\alpha}{\rm{argmin}} \left\{ \frac{1}{n}\sum_{i=1}^{m}||\bm{D}^{k}\alpha_i- {\textbf u}_i||^2+\lambda|\alpha_i|\right\}\\
\bm{D}^{k+1}&=&\underset{\bm{D}}{\rm{argmin}} \left\{\frac{1}{n}\sum_{i=1}^{m}||\bm{D}\alpha_i^{k+1}- {\textbf u}_i||^2+\lambda|\alpha_i|\right\}
\end{eqnarray}
As in \cite{Mairal:2009} we use a block-coordinate descent method \cite{chambolle:2005} for solving $\bm{D}$ and $\alpha_i$ iteratively. 

\section{Data selection and dictionary generation}
\label{section:setup}

In our previous work we applied learned dictionaries to denoise numerically generated gravitational waveforms from simulations of supernovae core-collapse and binary black hole mergers~\cite{Torres:2016} and to classify simulated glitches~\cite{Miquel:2019}. The data employed was in either case embedded in non-white Gaussian noise to simulate the background noise of advanced LIGO in its broadband configuration. This work takes a step further in our efforts by tackling the denoising problem with dictionaries employing real data, in the form of actual glitches from advanced LIGO's O1 data.

%
\begin{figure}[t]
 	{\includegraphics[width=0.4\textwidth]{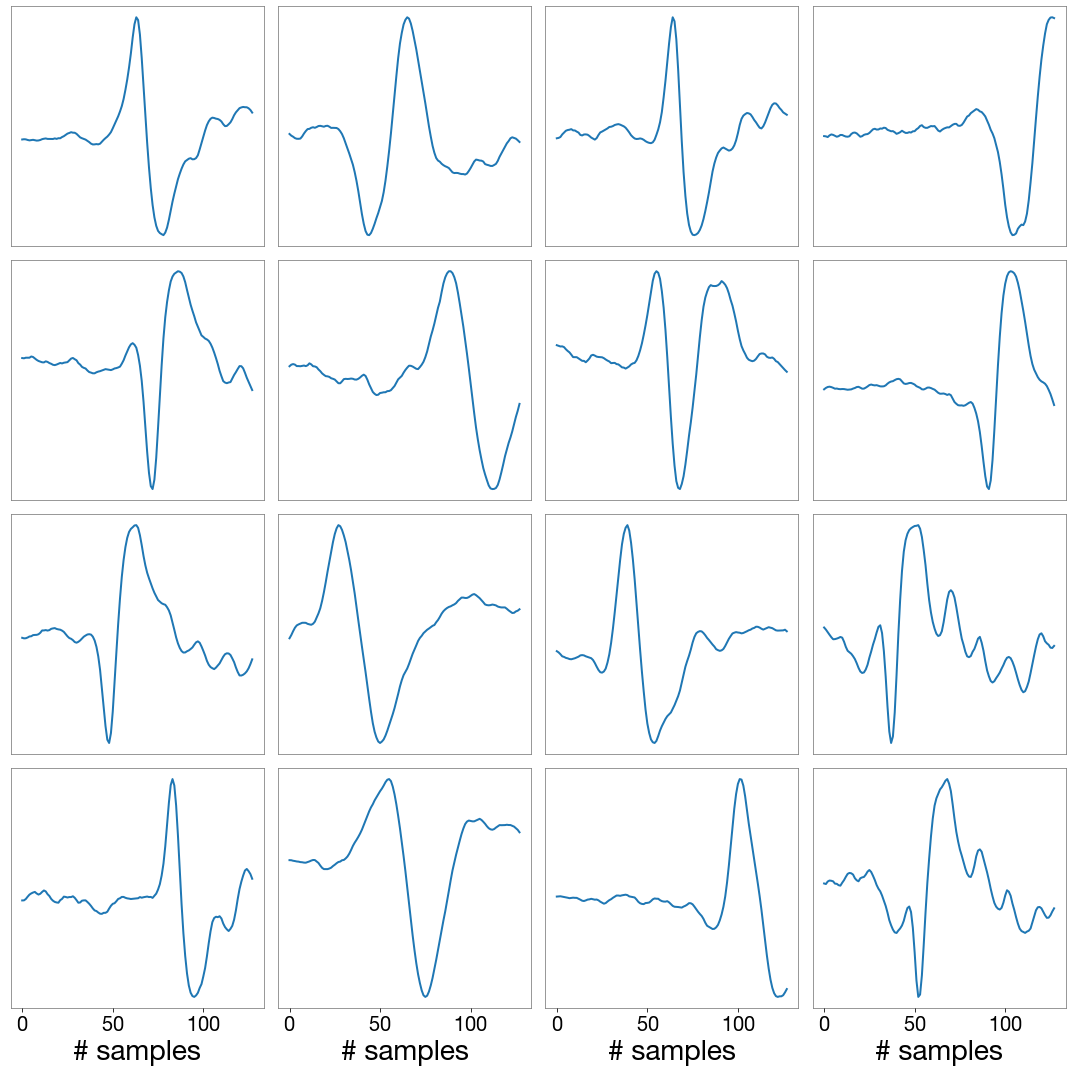}}
	 \caption{Example of a dictionary composed by a total of 192 atoms with 128 samples each. Only 16 atoms randomly selected are shown.}
 \label{fig:dict}
\end{figure}

We focus on blip glitches, the most common type of glitch found in the two LIGO detectors and whose origin remains mostly unknown~\cite{Cabero:2019}. Blip glitches, characterised by a duration of $\sim10$ ms and a frequency bandwidth of $\sim100$ Hz, have a distinctive tear-drop shape morphology when seen is a time-frequency (spectrogram) plot. By their intrinsic properties and recurring presence they can significantly reduce the sensitivity of searches for high-mass compact binary coalescences. In order to focus only in these noise transients, we apply a whitening procedure to the O1 data to remove all systematic sources of noise from the data, like the calibration and control signals that appear as lines in the spectrum, and also to flatten the data in frequency. The whitening algorithm uses the autoregressive (AR) model of~\cite{Cuoco:2001A, Cuoco:2001B}. The AR model employs 3000 coefficients estimated using 300 s of data at the beginning of the corresponding science segment of every glitch. This type of whitening has proved to be very robust and, as it is applied in the time domain, it is not affected by the typical border problems that appear in frequency-domain methods. 

To measure the accuracy of the glitch reconstruction and mitigation we employ two quantitative estimators. The first one is based on the time-frequency distribution of the power of the signal. We integrate the power spectrum for all frequencies for each temporal bin, and then we calculate the ratio between the maximum power and the mean power for all times. We will refer to this estimator as SNR:
\begin{eqnarray}
\label{eq:snr}
{\rm SNR}=\frac{\max (S(t))}{\overline{S(t)}}, \,\,\,S(t) = \int_{20}^{\frac{fs}{2}} S(t,f) \,df,
\end{eqnarray}
where ${S(t,f)}$ is the time-frequency representation of the data, and ${f_s}$ is the sampling frequency.  This SNR estimator is different from the one provided by the optimal filter, which is based in theoretical templates.
Our second estimator is called the ``whiteness''~\cite{Cuoco:2001B}
\begin{equation}
\label{eq:snr}
W = \frac{\exp(1/f_s\int ^{f_s/2}_{−f_s/2} \ln(P(f))\,df)}{1/f_s\int^{f_s/2}_{−f_s/2}P(f)\,df},
\end{equation}
where $P(f)$ is the power spectral density (PSD) of the data. The whiteness measures the spectral flatness. As the presence of glitches implies an increment of power with respect to a glitch-free background, if $P(f)$ is very peaky then $W\sim 0$, and if $P(f)$ is flat then $W=1$.

\subsection{Training}

We randomly select 100 blip glitches scattered in the data from advanced LIGO's first observing run. While this is not a big sample, it seems sufficient to assess the performance of learned dictionaries in removing glitches. The data corresponds to a window of 1 s centred at the GPS time of the glitch as provided by Gravity Spy~\cite{Zevin:2017}. The data is divided in two different sets; $85\%$ is used to train the dictionary while the remaining $15\%$ (which includes 16 blip glitches) is used to test the algorithm. The morphologies of all 16 glitches are presented in Appendix \ref{appendix}. The data is downsampled from their original 16384 Hz to 8192 Hz to speed up the algorithm and reduce the computational cost. 

The training process is performed as follows. After whitening all the data, we select the main glitch morphology using a window of 1024 samples around the GPS time of the glitch. Then, data from all glitches is aligned and organised in a matrix to build the initial dictionary. Next, we select 30000 random patches of a given length and start the block-coordinate descend method to obtain the trained dictionary. The length of the patches is the same as the atoms of the dictionary and, jointly with the number of atoms, is a hyperparameter of the model. 

We explore dictionaries formed by atoms of length in the range $[2^3,2^9]$. We also vary the number of atoms of each length to understand its possible effect on the results. Our study shows that a dictionary of 128 samples for atoms is a good choice. The number of atoms seems not to be very relevant as long as the dictionary is over-completed, i.e.~the number of atoms is larger than the length of the atoms.

One intrinsic difficulty of applying dictionary-reconstruction techniques to noise transients instead of to actual signals is that we lack of a ``clean'' signal to use as a model to the dictionary. Glitches have a random component due to the background. Even though the learning step has denoising capabilities, the block-coordinate descend method can have problems of convergence when the patches contain a large stochastic component. To improve the convergence of the learning step and the extraction results, we introduce an additional step between the whitening and the patch extraction. Namely, we use the rROF method to reduce the variance of the data used for training. This process results in smoother atoms and in a cleaner reconstruction of the glitch, which translates in a better separation between the background and the glitch morphology. An example of a dictionary is shown in Fig.~\ref{fig:dict}.

\subsection{Reconstruction}

Once the training step is complete, we use the resulting dictionary to extract the blip glitch from the background. As the length of the atoms is always shorter than the length of the test signals, we perform the reconstruction with a sliding window with an overlap of $n-4$ samples, where $n$ is the length of the atoms. The overlapped samples are averaged to obtain the final reconstruction.

In addition, we apply the ADMM algorithm in an iterative way. Starting with the original data $y$, we perform a reconstruction over the dictionary $u$. Then, this reconstructed signal is subtracted from the original data and the resulting residual is used as the new input. This procedure converges in the sense that in each iteration we subtract less signal from the background. Therefore, this iterative produce is applied until the differences between the residuals of consecutive iterations is less than a given tolerance. For most cases, a typical tolerance of $10^{-3}-10^{-4}$ is enough to produce good results.

\subsection{Regularization parameter search}
\label{section:parameter}

Reconstruction results heavily depend on the value of the regularization parameter $\lambda$. If its value is large, the relative weight of the regularisation term in Eq.~(\ref{eq:lasso}) is larger and more atoms are used. On the contrary, with a low value of $\lambda$ less atoms are used and more details of the signal (and noise) are recovered. In previous papers~\cite{Torres:2014, Torres:2016, Torres:2018}, we found the optimal value, i.e.~the one that produces the best results, comparing the denoised signal with the original one from GW catalogs from numerical relativity. In the present case, as there is not a true signal to compare with, we cannot determine the optimal $\lambda$ in the same way.

\begin{figure*}[t]
\centering
	{\includegraphics[width=0.28\textwidth]{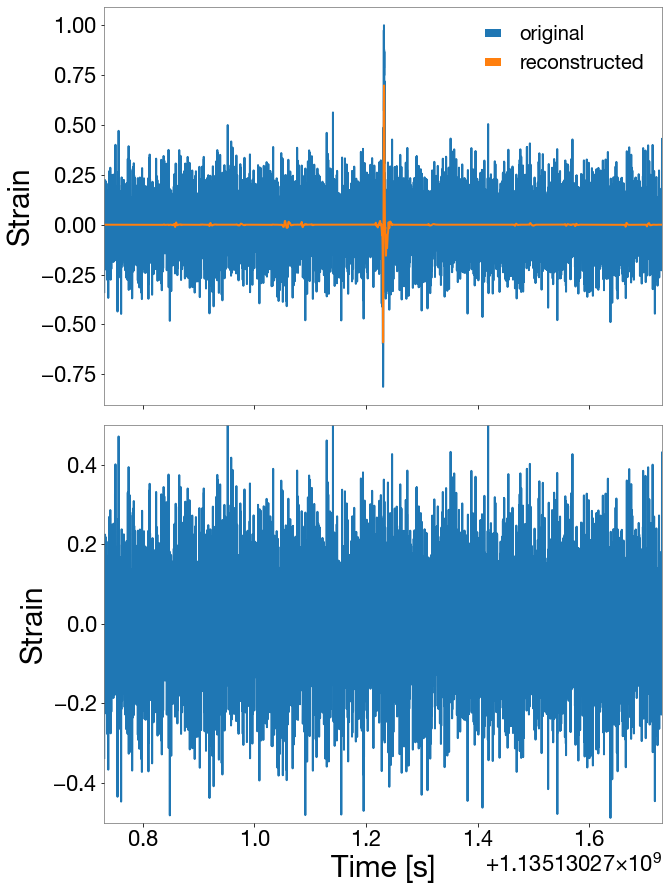}}
	{\includegraphics[width=0.28\textwidth]{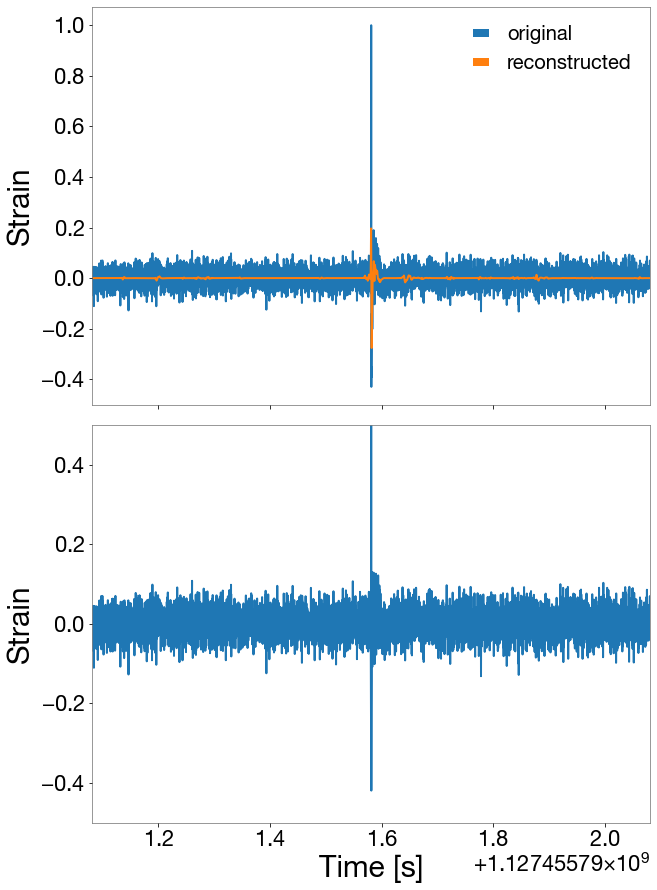}}
	{\includegraphics[width=0.28\textwidth]{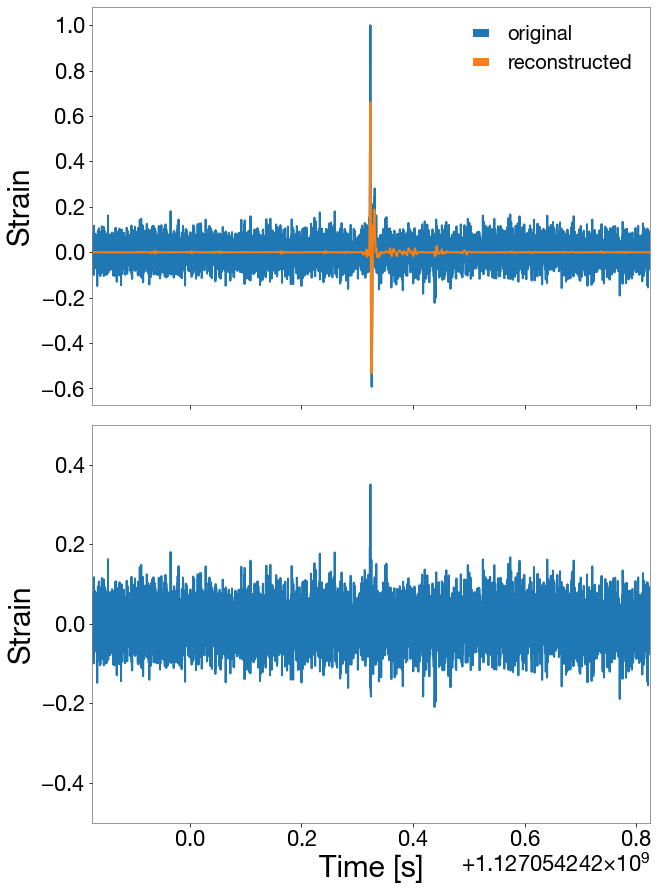}}
\caption{Time-series plots of three illustrative examples of blip glitches, corresponding to numbers 7, 9 and 15 from the test set. The upper panel shows the original signal (blue) with the reconstructed glitch (orange) superimposed using a dictionary of 192 atoms of 128 samples. The residual is shown in the bottom panel (Note that the interval of the vertical axes is smaller.).}
 \label{fig:recons_timeseries}
\end{figure*}
\begin{figure*}[t]
\centering
	{\includegraphics[width=0.28\textwidth]{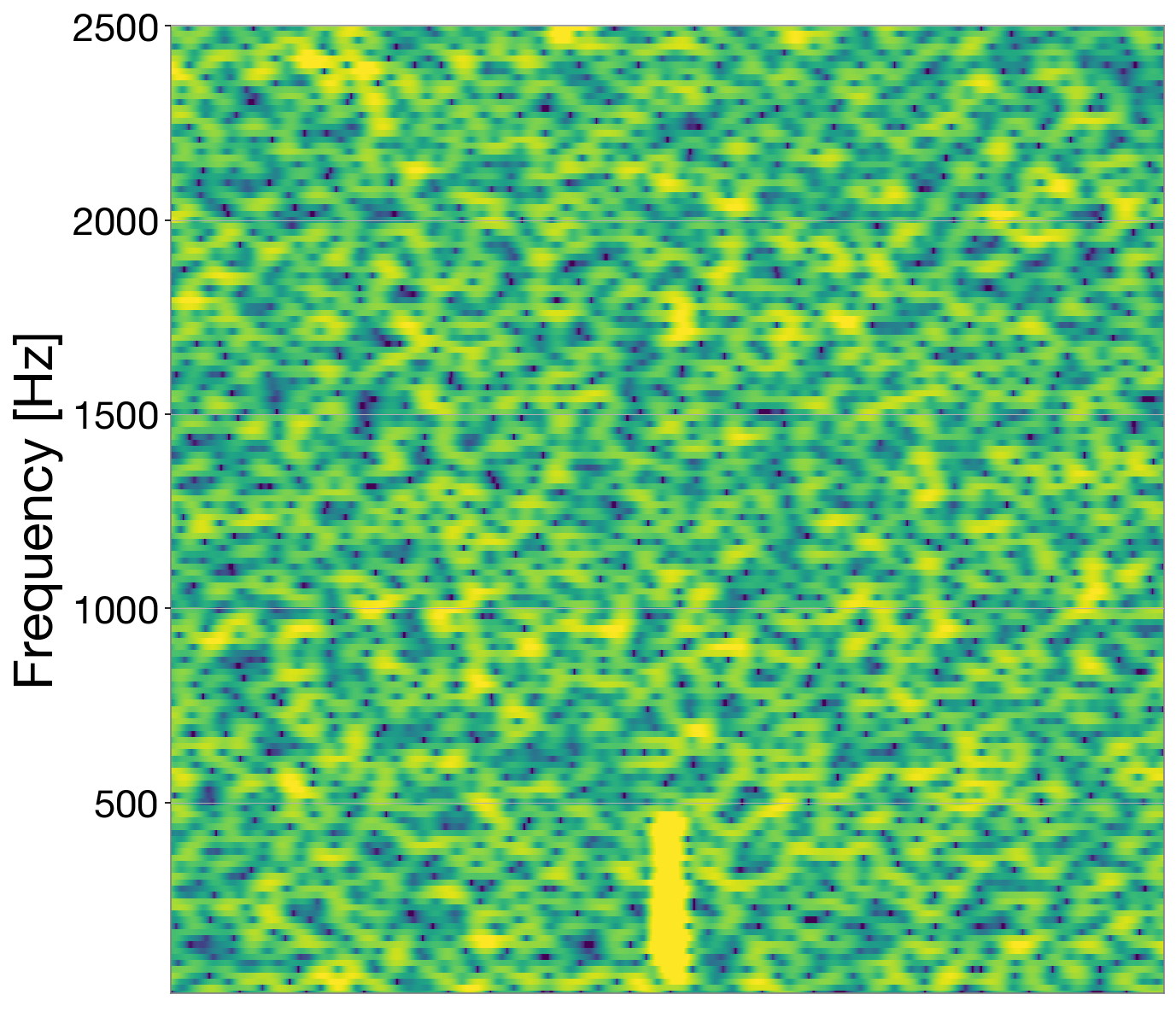}}
	{\includegraphics[width=0.28\textwidth]{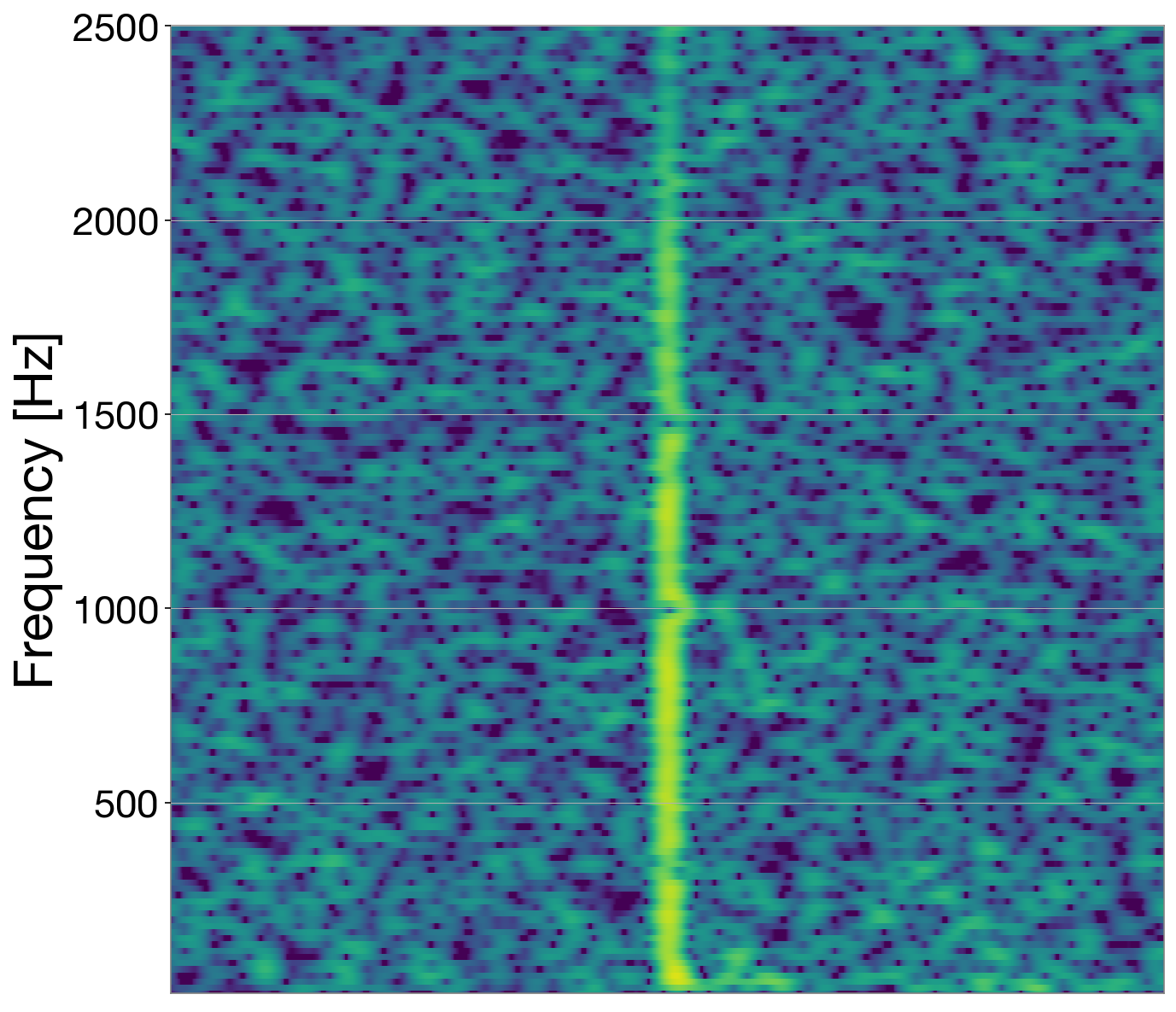}}
	{\includegraphics[width=0.28\textwidth]{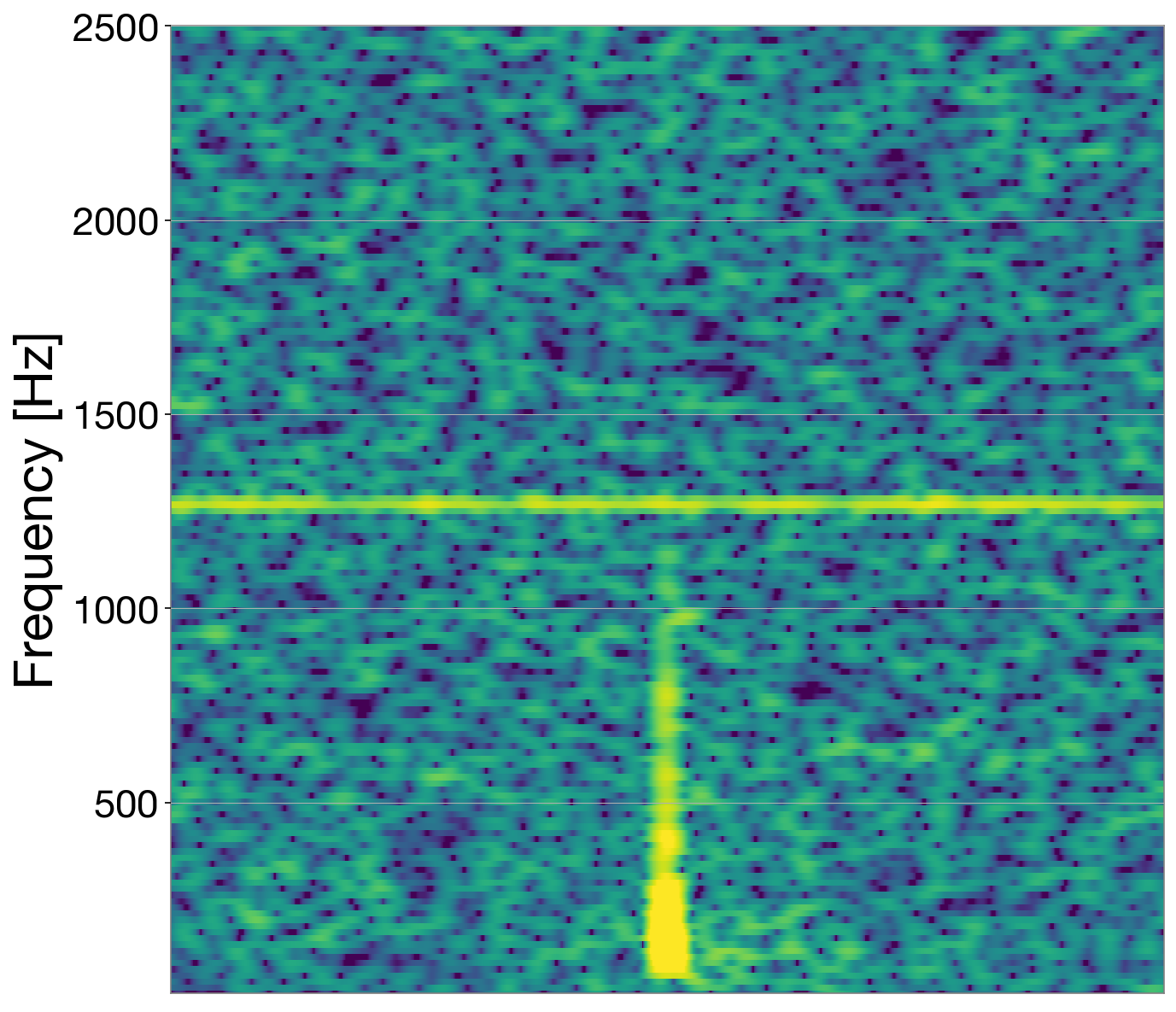}}\\
	{\includegraphics[width=0.28\textwidth]{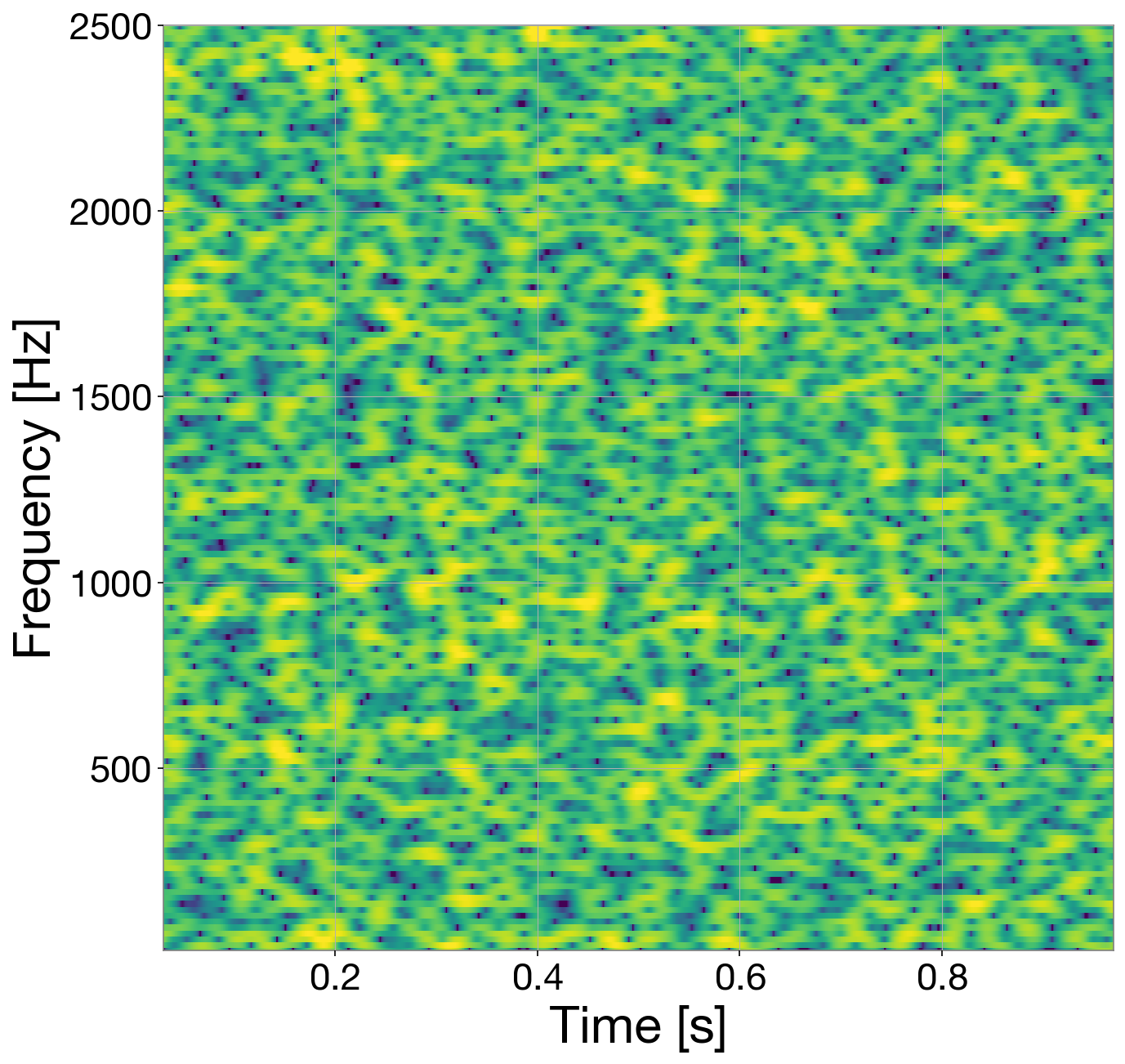}}
	{\includegraphics[width=0.28\textwidth]{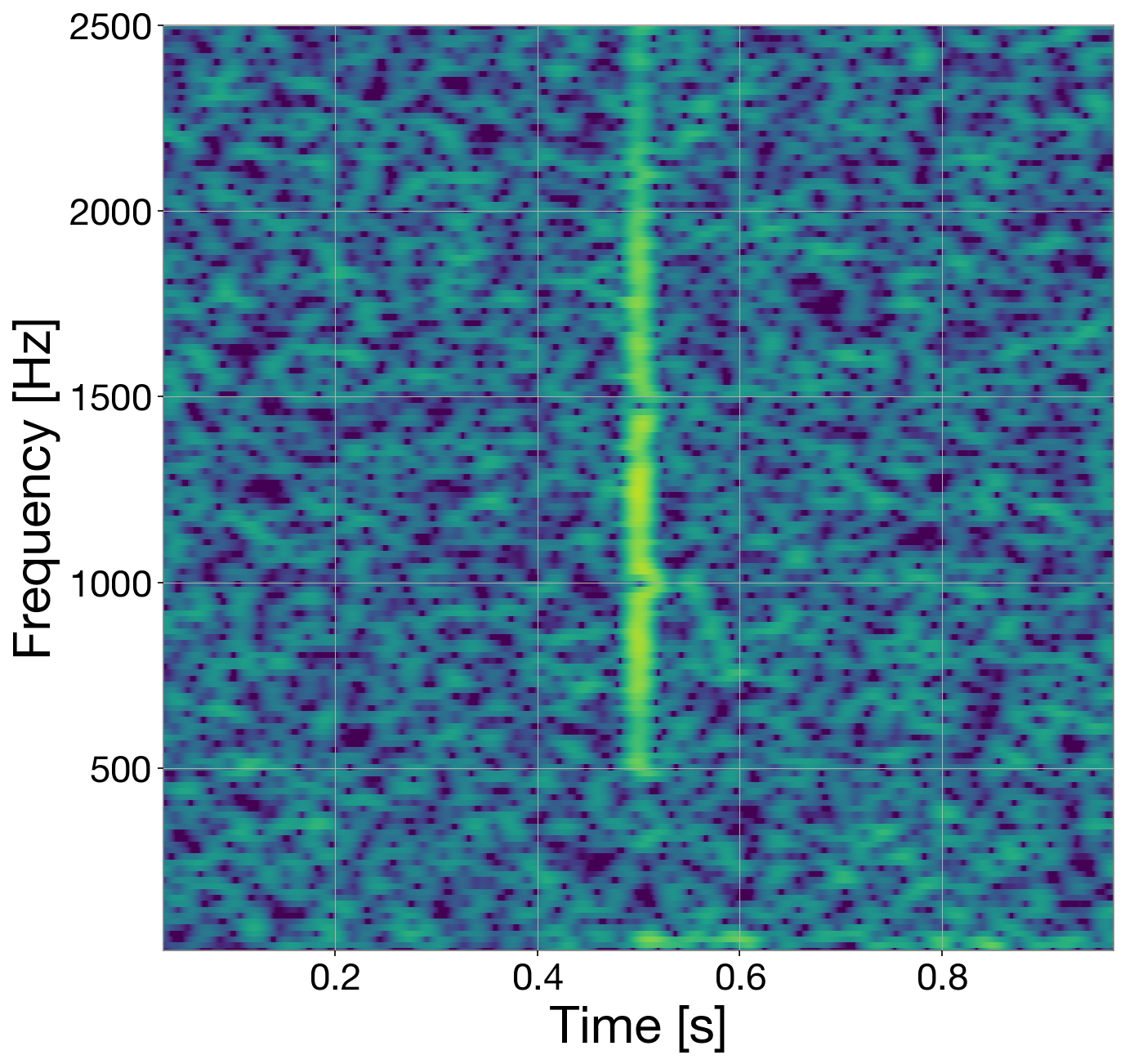}}
	{\includegraphics[width=0.28\textwidth]{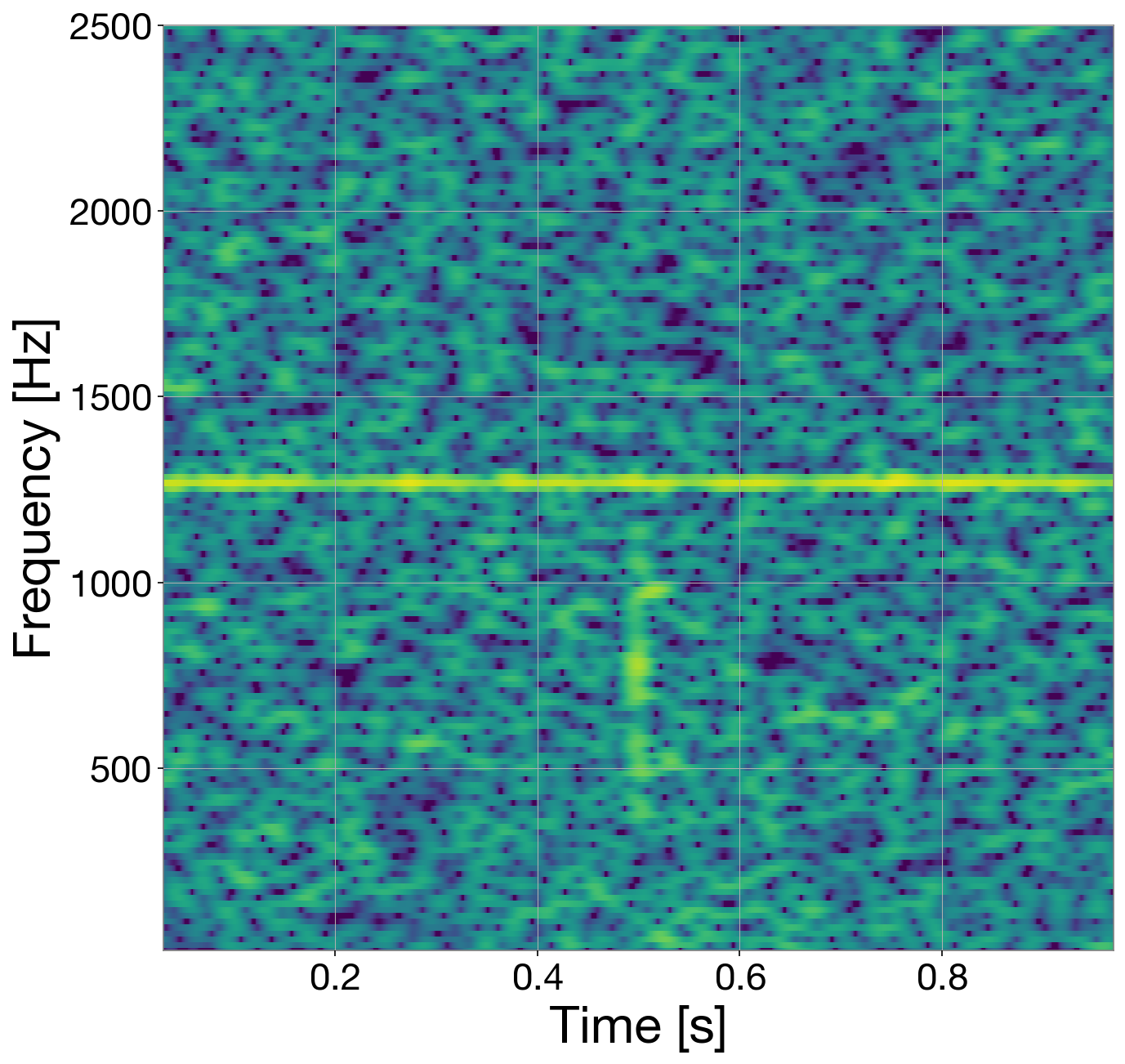}}
\caption{Time-frequency diagram of the same three examples in Fig.~\ref{fig:recons_timeseries}. The original data and the residuals are represented in upper and bottom panel respectively. }
 \label{fig:spect_single}
\end{figure*}

However, as our goal is to extract the glitch trying to keep the background unaltered, we design a different approach to obtain an optimal value of $\lambda$. One of the good properties of dictionary-learning methods we found in our previous studies is that the reconstruction returns zero when the data are very different from the dictionary. This statement is true for sufficiently large values of $\lambda$. As mentioned before, if the value is too low, the regularization term in Eq.~(\ref{eq:lasso}) becomes negligible and we would be solving essentially a least mean-squares fit, which in practice translates in a very oscillating reconstruction. If the value is very large, it is the fidelity term the one that becomes negligible, and the problem transforms in the minimization of the $\rm{L}_1$-norm of the vector $\alpha$, whose solution is the vector zero. Therefore, our goal is to find the first value of $\lambda$ that returns zeros in the part of the data dominated by the background but also produces a reconstruction for the glitch. The resulting reconstruction will be zero on the window border and will avoid discontinuities. We select a small window at the beginning of the data window where we are sure that the data is mostly dominated by the background. After that, a bisection algorithm tries to find the largest value that returns a non-zero reconstruction. We will refer to this value as $\lambda_{\rm{min}}$.

\section{Results}
\label{section:results}

\subsection{Blip glitch subtraction with a single dictionary}
\label{subsection:single_dictionary}

%
\begin{figure*}[t]
\centering
	{\includegraphics[width=0.28\textwidth]{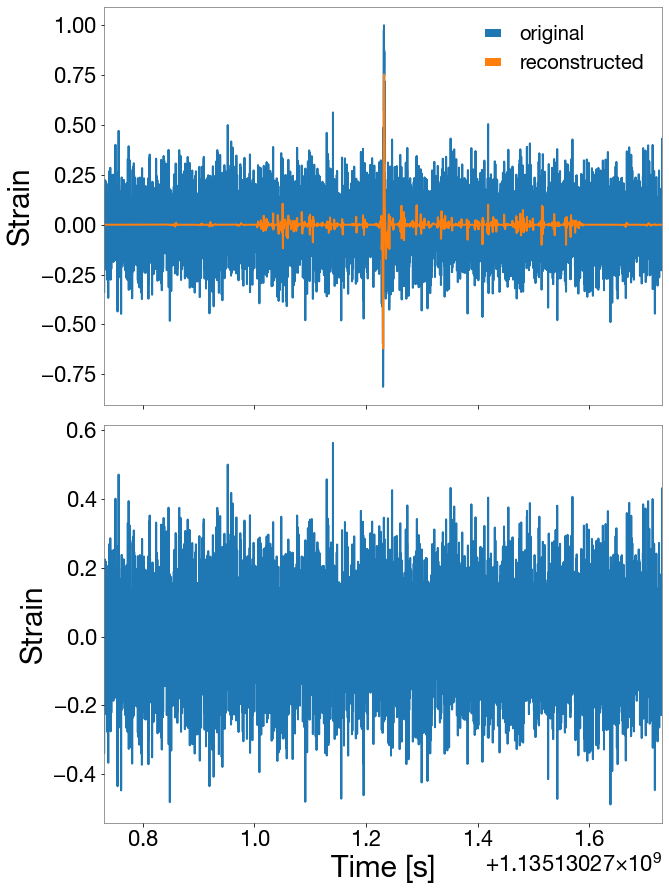}}
	{\includegraphics[width=0.28\textwidth]{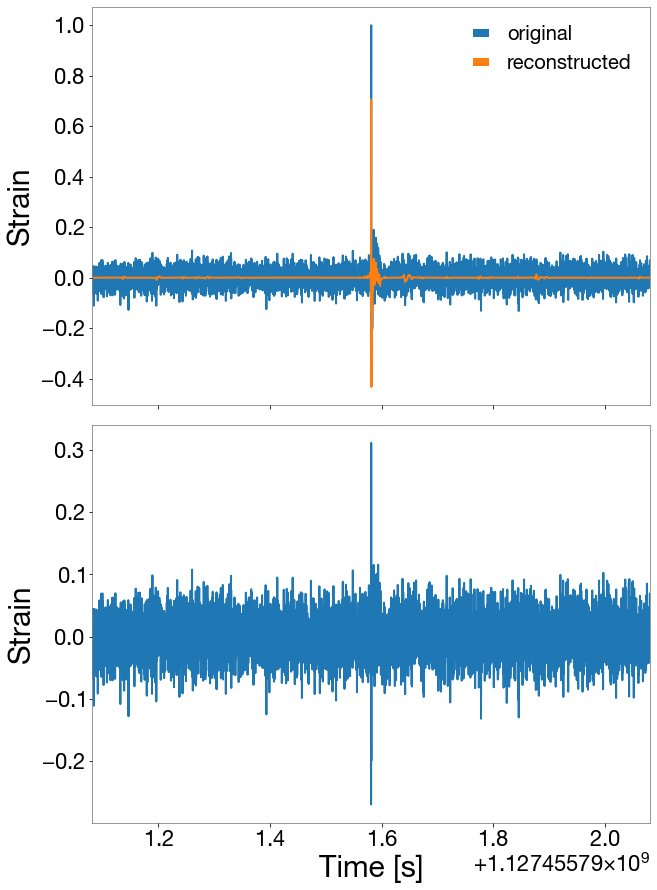}}
	{\includegraphics[width=0.28\textwidth]{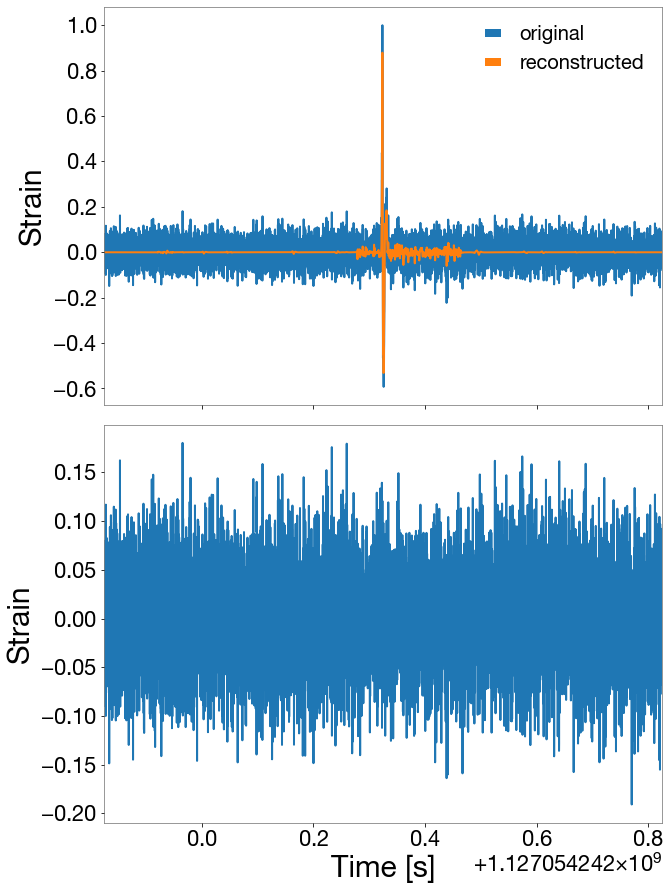}}
\caption{Time-series plot of the same three examples of Fig.~\ref{fig:recons_timeseries} when the reconstruction is done combining dictionaries. The original data and the residuals are shown in the top and bottom panels, respectively. }
 \label{fig:multi_recons_timeseries}
\end{figure*}
\begin{figure*}[t]
\centering
	{\includegraphics[width=0.28\textwidth]{./Figures/spect_single_7_original.png}}
	{\includegraphics[width=0.28\textwidth]{./Figures/spect_single_9_original.png}}
	{\includegraphics[width=0.28\textwidth]{./Figures/spect_single_15_original.png}}\\
	{\includegraphics[width=0.28\textwidth]{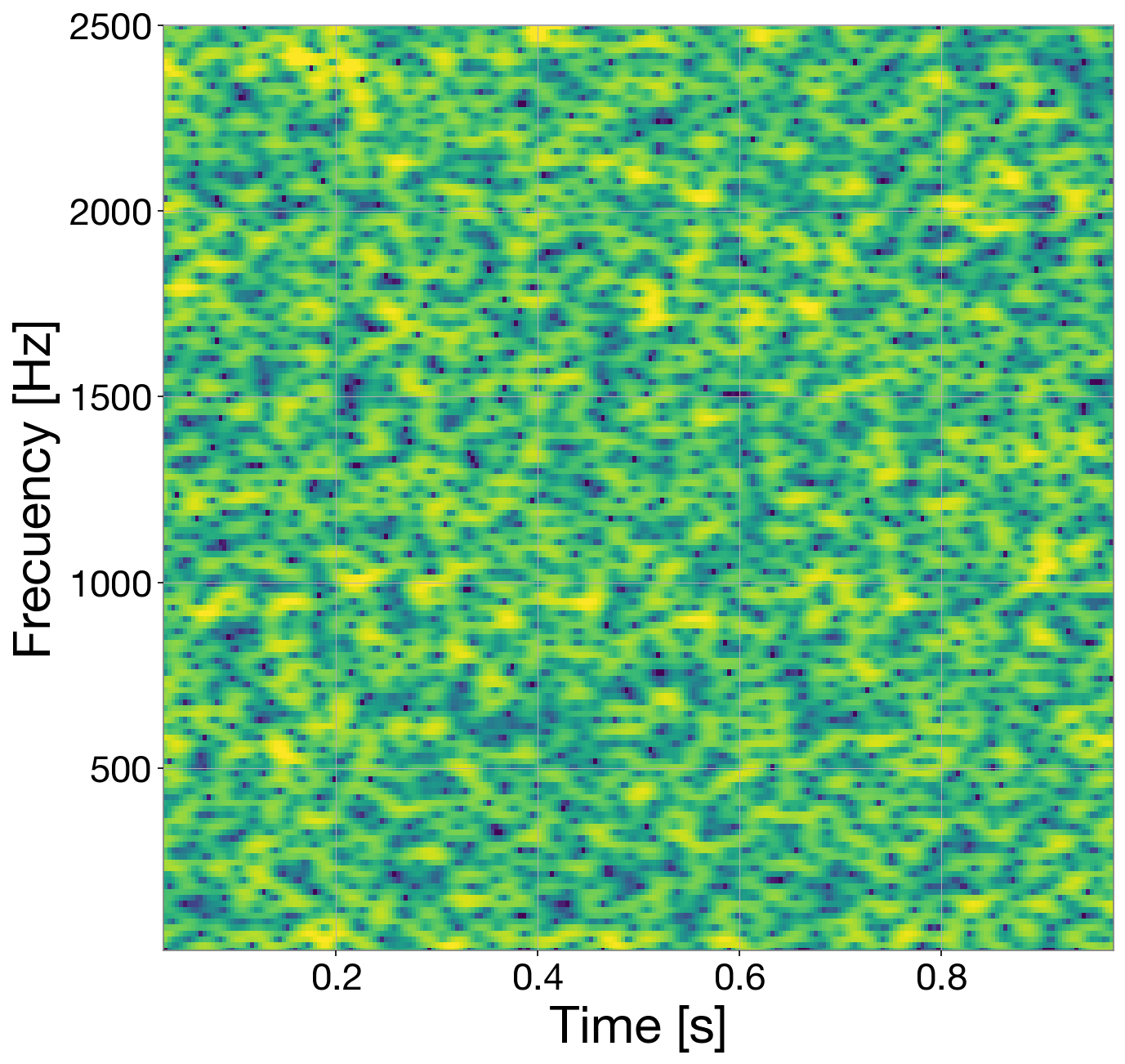}}
	{\includegraphics[width=0.28\textwidth]{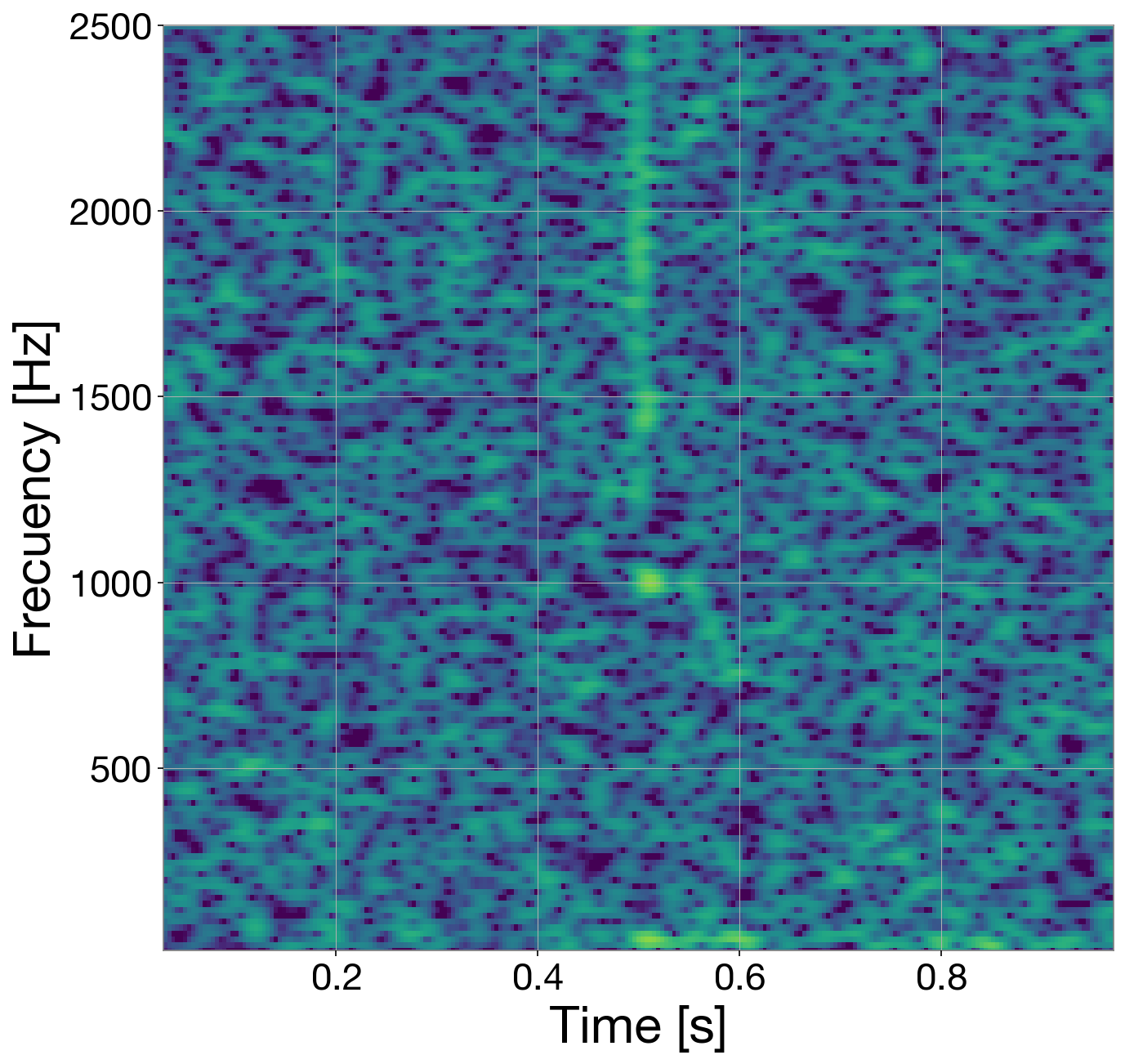}}
	{\includegraphics[width=0.28\textwidth]{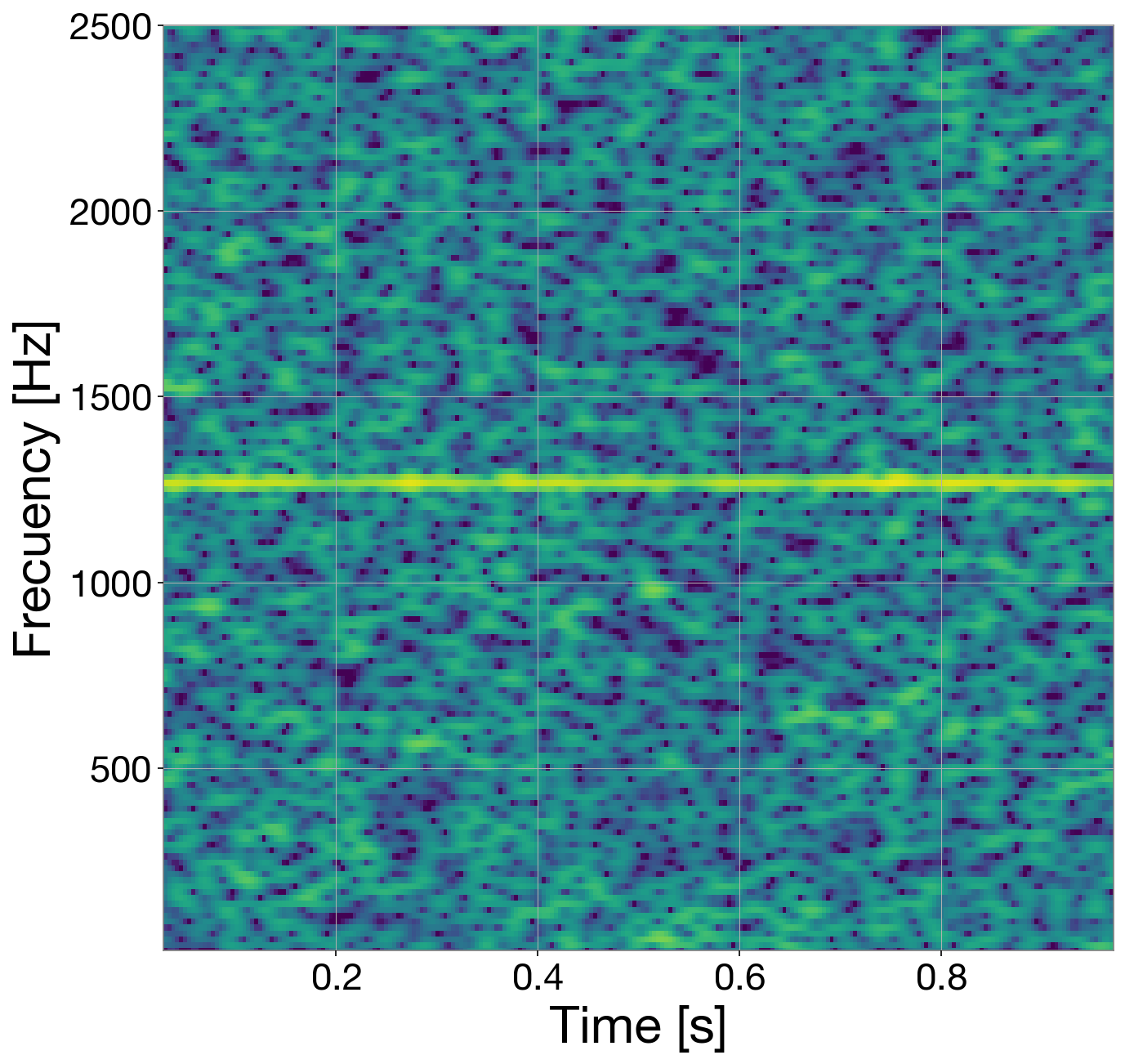}}
\caption{Time-frequency diagram of the same three examples of Fig.~\ref{fig:recons_timeseries} when using combined dictionaries.}
\label{fig:spect_multi}
\end{figure*}

We start applying dictionaries of different length and number of atoms to the 16 blip glitches we use as tests. The bisection procedure introduced in Section \ref{section:parameter} is used to find the lower value of $\lambda$ that returns zeros at the beginning of the data stream. The reconstructed glitch is then subtracted from the data to obtain a cleaner background.

As mentioned previously, we explore the results of using dictionaries formed by atoms of length inside the range $[2^3,2^9]$. All options are able to reconstruct the glitches and reduce their impact significantly. However, we observe slightly different results depending of the length of the atoms. In particular, dictionaries with shorter atom length are able to extract more high frequency features than those with larger atom length, but at the expense of leaving more energy at lower and middle frequencies. We have also explored the effect of the number of atoms. We find that once the number of atoms is sufficient large, above $\sim2.5$ times the length of the atoms, there is no appreciable improvement in the results.

Fig.~\ref{fig:recons_timeseries} displays the time-series of three illustrative examples of our sample of blip glitches reconstructed with a dictionary of 192 atoms, each with a length of 128 samples. Correspondingly, Fig.~\ref{fig:spect_single} shows the time-frequency diagram (spectrogram) of the same data. In all these examples (and in the whole set of glitches of our sample) the amplitude of the reconstruction is almost zero in the parts of the data stream that contain only background, and only the parts of the signals with larger amplitude are reconstructed. The examples in Fig.~\ref{fig:recons_timeseries}reveal that the algorithm is able to reconstruct all blips present in the data. The subtraction residuals (bottom panels of Fig.~\ref{fig:recons_timeseries}) show that the impact of the glitches is significantly reduced in all cases. However, some non-negligible part of the glitch still remains in the data (see middle and right panels).

Let us now focus on the spectrograms shown in Fig.~\ref{fig:spect_single}. The upper panels display the original data (after the whitening procedure) and the lower panels show the reconstructed spectrograms obtained when using a single dictionary of 192 atoms, each with a length of 128 samples. The three blips chosen to illustrate our procedure show several features that make them interesting cases of study. The left panel shows a blip with the typical tear-drop shape morphology. The middle panel displays a blip glitch with a strong contribution at high frequencies. Finally, a strong spectral line around $\sim$1400 Hz is well visible in the right panel and it is simultaneous to the occurrence of the blip glitch. The spectrograms of the reconstructed data confirm the analysis of the time-series plots. The power of the glitches is reduced for the low and middle frequencies (up to $\sim500$ Hz) while in other parts of the spectrogram both the signal and the background remain unaltered. This effect is clearly visible in the right plot of the lower panel of Fig.~\ref{fig:spect_single} which shows that the prominent spectral line is still present after the reconstruction. This is both a good and a bad feature of the method. On the one hand, it is a good result because, as the morphology of the line is totally different to that of the blip glitch, the dictionary does not reconstruct it; that would be a nice feature of the method in the case of a coincidence (both temporal and in frequency) of a blip glitch with an actual GW signal. On the other hand, it is a bad result because, as spectral lines are another source of noise, they require additional techniques to mitigate their impact. 

\begin{table}
\begin{center}
\begin{tabular}{ccccccc}
Test \# & $\rm{SNR_o} $& $\rm{SNR_{single}}$ & $\rm{SNR_{multi}}$ &$\rm{W_o}$ &$\rm{W_{single}}$&$\rm{W_{multi}}$ \\ 

\hline
 1 & 5.3 & 1.3 & 1.3 & 0.99 & 0.99 & 0.99 \\ 
 2 & 5.2 & 2.4 & 1.3 & 0.94 & 0.94 & 0.93 \\ 
 3 & 9.2 & 1.3 & 1.3 & 0.99 & 0.99 & 0.99 \\ 
 4 & 37.1 & 10.2 & 1.5 & 0.84 & 0.92 & 0.97 \\ 
 5 & 18.1 & 2.3 & 1.7 & 0.99 & 0.98 & 0.98 \\ 
 6 & 13.5 & 3.8 & 1.7 & 0.98 & 0.98 & 0.98 \\ 
 7 & 4.1 & 1.3 & 1.3 & 0.98 & 0.98 & 0.98 \\ 
 8 & 6.3 & 1.3 & 1.3 & 0.96 & 0.97 & 0.97 \\ 
 9 & 13.4 & 9.8 & 2.9 & 0.98 & 0.98 & 0.98 \\ 
10 & 8.7 & 3.0 & 1.7 & 1.00 & 0.99 & 0.99 \\ 
11 & 7.3 & 1.3 & 1.3 & 0.99 & 0.99 & 0.99 \\ 
12 & 5.8 & 1.3 & 1.2 & 0.99 & 1.00 & 1.00 \\ 
13 & 4.5 & 1.8 & 1.3 & 0.99 & 0.99 & 0.99 \\ 
14 & 14.2 & 1.2 & 1.2 & 0.99 & 0.99 & 0.99 \\ 
15 & 15.2 & 2.2 & 1.3 & 0.88 & 0.88 & 0.89 \\ 
16 & 6.2 & 1.3 & 1.3 & 0.99 & 0.99 & 0.99 \\ 
\hline
\end{tabular}
\end{center}
\label{tab:metrics}
\caption{Quantitative assessment of our deglitching procedures. The columns report the values of our estimators, SNR and W, for the data containing the original noise transients (subindex `o') and for the residuals after deglitching, and both for a single dictionary (subindex `single') and for multiple dictionaries (subindex `multi').}
\end{table}

The most obvious conclusion from the spectrograms is that the dictionary has difficulties in reducing the high-frequency content of the glitches. In the next subsection we discuss how to improve the performance at high frequencies and present quantitative estimates, using the SNR and W metrics, of our two deglitching procedures.

\subsection{Combining dictionaries}
\label{subsec:combination}

We turn next to describe the results obtained when using a combination of different dictionaries to improve the results at high frequencies (above $\sim$500 Hz). This combination is a natural extension of our original algorithm based on a single dictionary. Now the algorithm reads as follows: first we perform the reconstruction in the same way than before using the largest dictionary (i.e.,~192 atoms of 128 samples length each). Then, a second dictionary is applied to the residual obtained from the first one. However, we do not use $\lambda_{\rm{min}}$ for this second dictionary. As our goal is to improve the results at high frequencies, we use a slightly lower value of $\lambda$, around $85\%$ less. To keep the background from being affected due to this lower value of $\lambda$ we restrict the reconstruction to a window which contains the most significant part of the blip. This window is determined by the reconstruction using the first dictionary, selecting the part of the signal which is not zero.

As an example we discuss results using a combination of two dictionaries, one formed by 192 atoms of 128 samples and another one comprising 40 atoms of only 16 samples. The results are presented in Figs.~\ref{fig:multi_recons_timeseries} and~\ref{fig:spect_multi}. The  comparison of the time-series plots of Fig.~\ref{fig:multi_recons_timeseries} and Fig.~\ref{fig:recons_timeseries} does not show an obvious improvement when two dictionaries are used instead of one. The inspection  of Fig.~\ref{fig:multi_recons_timeseries} reveals that two of the peaks at the maximum of the glitch disappear in the left and right panels while part of the peak of the glitch in the center panel is also reduced. Comparing the spectrograms (i.e.~Figs.~\ref{fig:spect_single} and \ref{fig:spect_multi}) yields more meaningful information. One can observe that the high-frequency contribution that remains from the blip reconstruction with a single dictionary (up to $\sim 1000$ Hz) is further reduced by using a second, smaller dictionary. In addition, the background is not significantly perturbed. This also holds when the first dictionary is able to extract the blip glitch completely, as shown in the left panel of Fig.~\ref{fig:spect_multi}.

We have also analyzed the results for lower values of $\lambda$ than the $0.85\lambda_{\rm{min}}$ value used in this example. We find that the high-frequency component of the glitch can be reduced even more. However, at low and middle frequencies the background is affected and the spectrogram shows a significant reduction of power in the glitch there. Our tests indicate that a value of $\lambda$ around $0.85\lambda{\rm{min}}$ yields a good tradeoff.

Table~I reports the two metrics, SNR and W, for all test cases shown in Figs.~\ref{fig:recons_timeseries}-\ref{fig:spect_multi}. The values of SNR and W for the original signals are shown in columns 2 and 5, respectively. The comparison using a single dictionary (columns 3 and 6) indicates that the dictionary is able to reduce the SNR significantly. The values of the whiteness increase sightly as expected. Note that as the data is whitened before the reconstruction, most of the original values of W are already close to one. For our procedure combining two dictionaries (columns 4 and 7 in Table~I) the estimators show that in those cases where the first denoising does not reduce the glitch completely, the second dictionary is able to improve the results. In addition, in cases where the first dictionary already shows good performance and the SNR is reduced significantly, the addition of a second dictionary barely modifies the results. As a summary we conclude that, in general, glitch denoising with multiple dictionaries is a convenient strategy. For our test set of 16 blip glitches it reduces the SNR by a factor of $\sim11$ on average, while with a single dictionary the average reduction is $\sim7.5$, with a negligible increment in computational cost.

\subsection{Deglitching of GW170817}
\label{GW170817}

\begin{figure}[t]
\centering
{\includegraphics[width=0.45\textwidth]{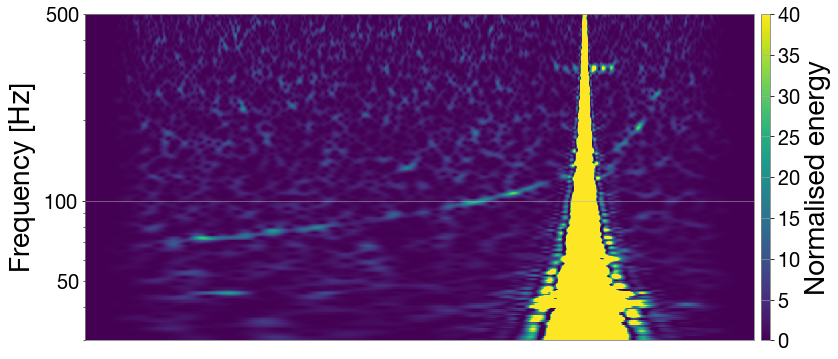}}
	{\includegraphics[width=0.45\textwidth]{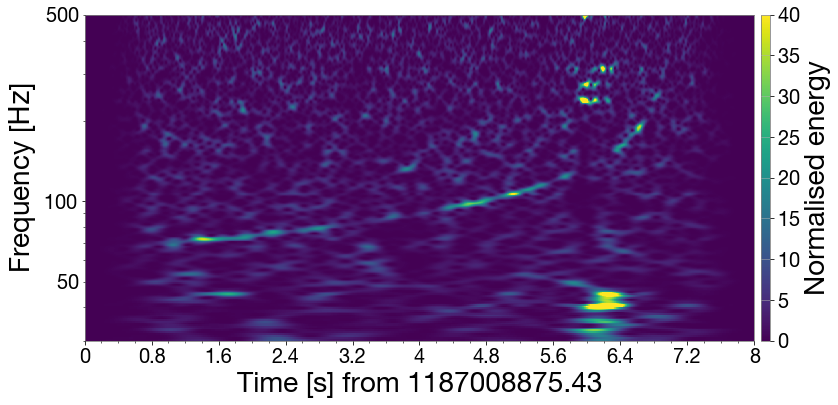}} \\

	\caption{Time-frequency diagram of 8 seconds of data corresponding to the GW170817 signal. The top panel shows the original data from LIGO-Livingston. The bottom panel displays the data after subtracting the glitch using a single blip-trained dictionary with 256 samples.}
\label{fig:GW170817}
\end{figure}

In~\cite{GW170817} the LIGO-Virgo Collaboration reported the first detection of GWs from a binary neutron star inspiral, GW170817. About $1.1$s before the coalescence time of GW170817, a short instrumental  noise  transient appeared in the LIGO-Livingston detector (see Fig.~\ref{fig:GW170817}, upper panel). The glitch was modeled with a time-frequency wavelet reconstruction and subtracted from the data, as shown in Fig.~2 of~\cite{GW170817}.

In this section we evaluate the use of learned dictionaries to deglitch the noise transient appearing in GW170817. The results are plotted in Fig.~\ref{fig:GW170817}. We note that the spectrograms in this figure look different to those shown in the previous sections. This is because we use the routines of the Q-transform included in the GWpy libraries~\cite{GWpy}  in order to obtain similar plots to those reported in~\cite{GW170817} to facilitate the comparison. It is worth stressing that the shape of the GW170817 glitch is not the same as that of the blip glitches we use to train our dictionaries. Therefore, these are not specifically tailored to deglitch the particular noise transient affecting GW170817. Nevertheless, this example still provides an excellent test to assess the capabilities of our trained dictionaries to reconstruct other types of glitches. In addition, the presence of the binary neutron star signal allows us to analyse if it is affected by the deglitching procedure.

The bottom panel of Fig.~\ref{fig:GW170817} shows the results after applying one single dictionary of 256 samples, which is the dictionary that produces the best results in terms of reducing the contribution of both high and low frequencies. This is expected  because the GW170817 glitch lasted significantly longer than the test cases discussed before. This figure shows that for the most part the glitch is removed from the data and, at the same time, the actual chirp signal from the inspiraling neutron stars behind the glitch is recovered almost completely. Nevertheless, part of the glitch at frequencies of $\sim 400$ Hz and below $\sim 50$ Hz remains visible after the first reconstruction. Subsequently applying a second dictionary composed by atoms of 16 samples does not significantly improve the results of the first dictionary in this example. This may be related to the fact that our trained dictionaries are not specifically customized to the morphological type of the GW170817 glitch. We note that these results could potentially improve if we trained our dictionaries with a larger set of glitch morphologies, instead of only using blips. We plan to investigate this possibility in the future.

\section{Summary}
\label{section:summary}

We have investigated the application of learned dictionaries to mitigate the effect of noise transients in the data of GW detectors. Although the data show the presence of many families of glitches, each with a different morphology and time-frequency shape, we have focused on ``blip" glitches because they are the most common type of glitches found in the twin LIGO detectors and their waveforms are easy to identify. This paper is the first step in our ongoing program to automatically classify and subtract all families of glitches employing variational methods.

Our approach combines two different variational techniques based on the $L_1$ norm, namely the Rudin-Osher-Fatemi method~\cite{Rudin:1992} and a Dictionary Learning method~\cite{Mairal:2009}. We have randomly selected 100 blip glitches scattered in the data from advanced LIGO's O1 run. The data corresponds to a window of 1 s centred at the GPS time of each glitch as provided by Gravity Spy~\cite{Zevin:2017}. $85 \%$ of the glitches have been used to train the dictionary while the other $15 \%$ have been employed as examples to test the performance of the algorithm. The test set has included 16 blip glitches. In our approach we have incorporated a regularized ROF denoising step before the training step to obtain a smooth dictionary, which has proved to be more effective to model and subtract the glitches from the data.

The determination of a good value of the regularization parameter $\lambda$ is not a trivial task. In this paper we have deviated from our previous works where the optimal value of $\lambda$ could be determined by comparison with an analytical or numerical template~\cite{Torres:2014, Torres:2016, Torres:2018, Miquel:2019}. To preserve the background as unaltered as posible, we find the first value of $\lambda$ that only produces a reconstruction of the glitch and returns zeros for the surrounding background. This procedure has the advantage that $\lambda$ is calculated only from the data and does not require any systematic search using templates. Our results have shown that this approach is valid to model and subtract most of the contribution of the blip glitches in all cases analyzed. They also have revealed, however, that the high-frequency component of the blips (above $\sim 500$ Hz) is not completely removed. This issue has been ameliorated by using a combination of dictionaries with different atom length, one much shorter than the other. This seems to be the best strategy to mitigate the blips at all frequencies. We have also shown that our approach yields satisfactory results when applied to the GW170817 glitch~\cite{GW170817} despite this transient noise feature is not of the blip glitch class.  Learned dictionaries trained with a different set of glitches can remove the GW170817 glitch from the data without affecting the actual signal from the binary neutron star inspiral. Potentially, this result could be further improved by training the dictionaries with larger  datasets accounting for other types of glitches.

In order to eventually employ this approach in a low-latency pipeline in actual GW detectors, it would be necessary to improve the computational performance of the method.  The current version of the code is implemented in MATLAB and not optimized. On average, the amount of time to process 1 s of data sampled at 8192 Hz is $\sim1.5$ s using a MacBook Pro with a 2,3 GHz Intel Core i5 processor and 16 Gb of RAM memory using 4 cores. We note, however, that the reconstruction can be trivially parallelised since each segment can be processed independently of the others. Therefore, there is still a considerable margin for improvement in execution time, developing an optimized and compiled version of the code. In the near future, we plan to extend the work initiated in \cite{Miquel:2019} to account for other families of glitches and use classification algorithms as a previous step to the glitch subtraction procedure presented in this work. 

\section*{Acknowledgements}

Work supported by the Spanish Agencia Estatal de Investigaci\'on (grant PGC2018-095984-B-I00), by the Generalitat Valenciana (PROMETEO/2019/071), by the European Union’s Horizon 2020 RISE programme H2020-MSCA-RISE-2017 Grant No.~FunFiCO-777740, and by the European Cooperation in Science and Technology through COST Action CA17137.

\appendix
\section {Properties of the test set}
\label{appendix}

\begin{figure*}[t]
\centering
	{\includegraphics[width=0.22\textwidth]{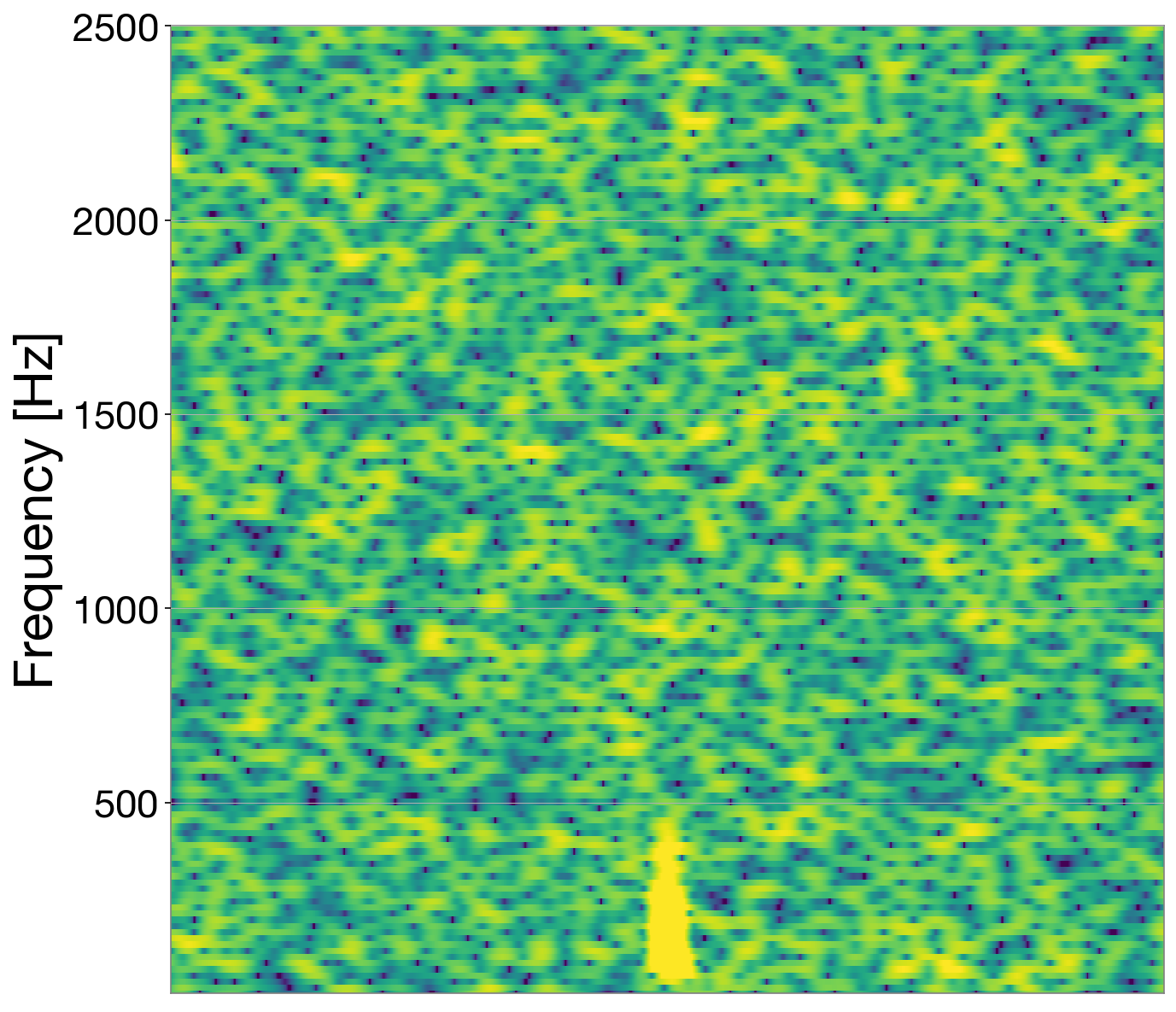}}
	{\includegraphics[width=0.22\textwidth]{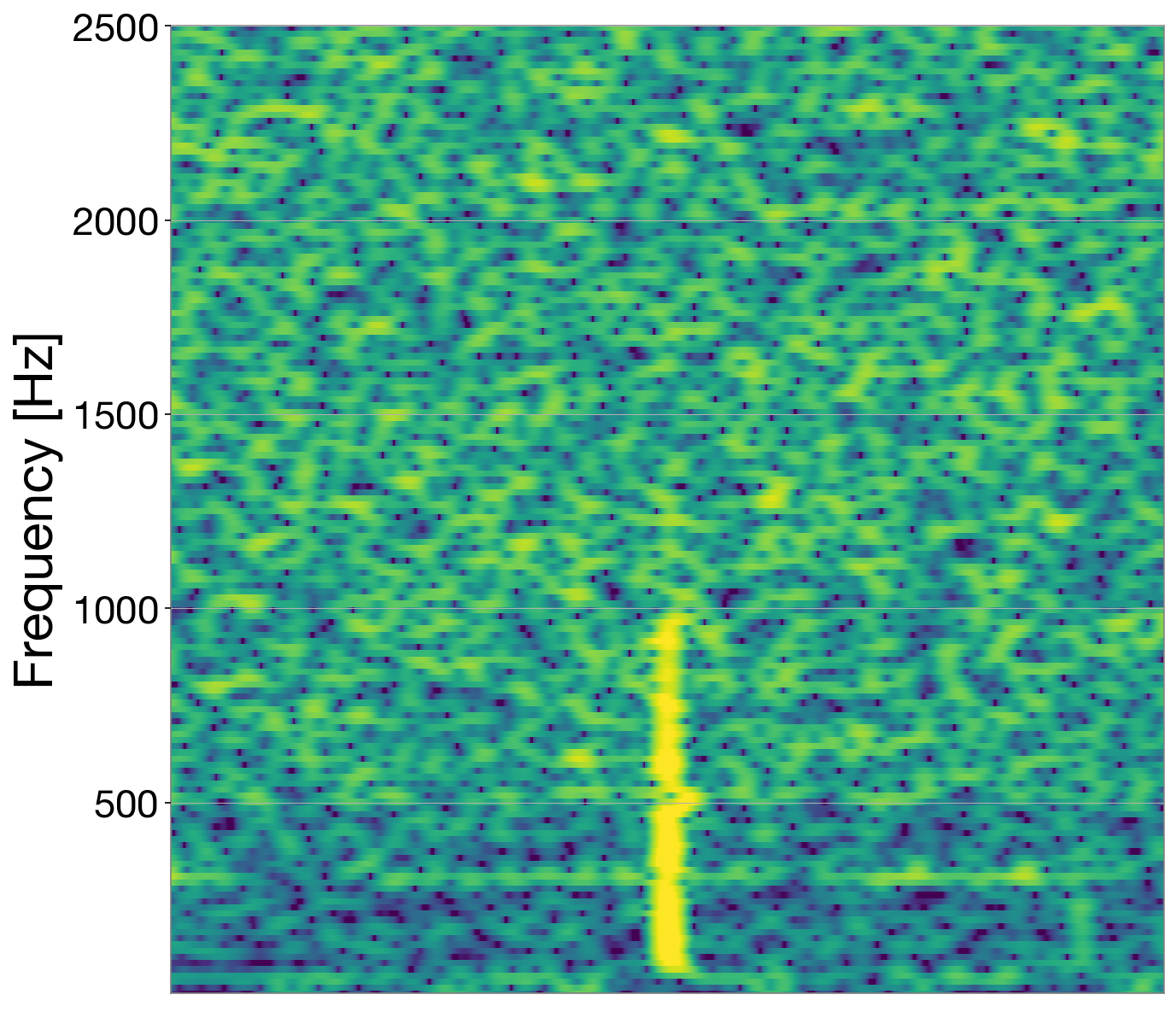}}
	{\includegraphics[width=0.22\textwidth]{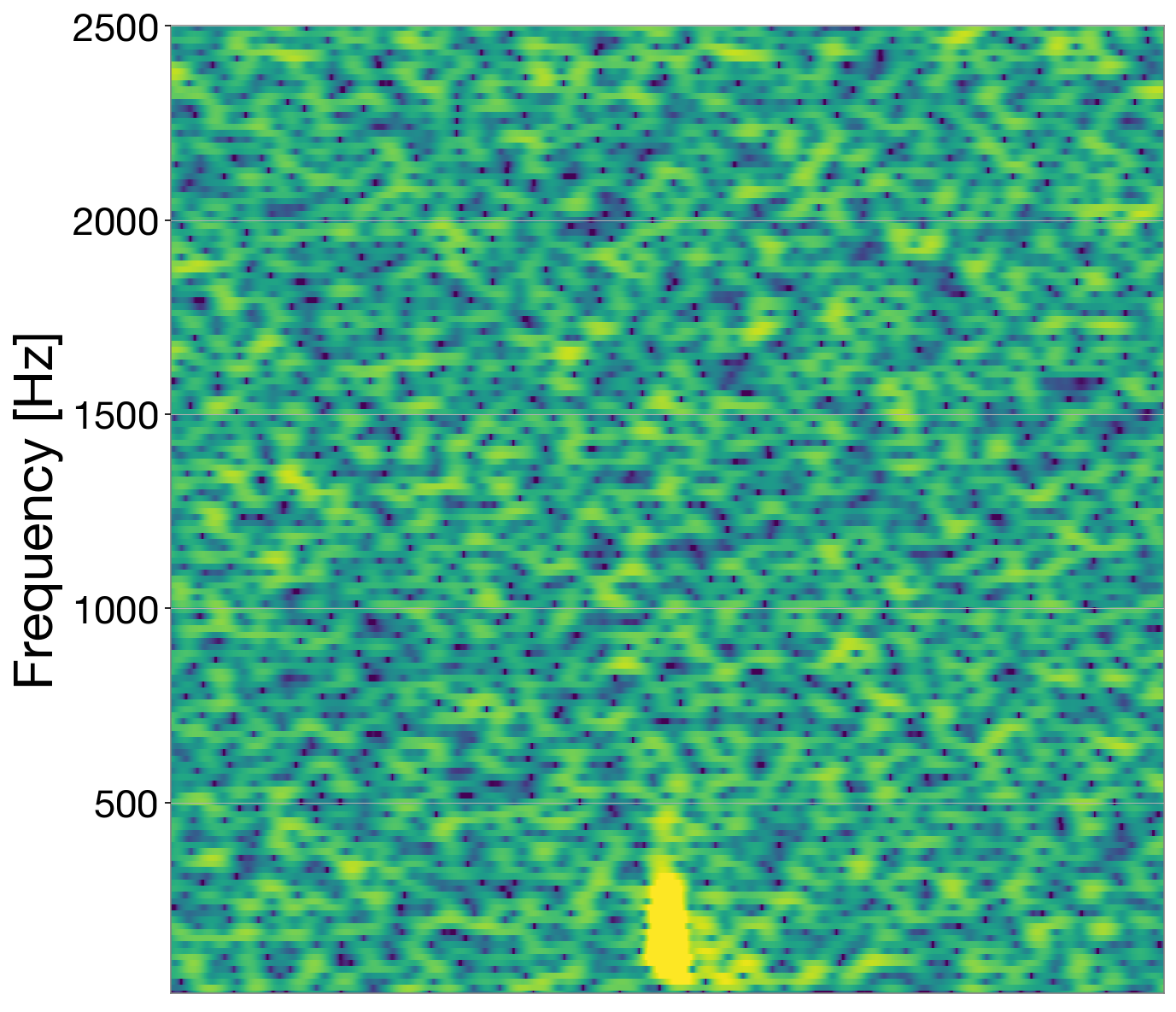}}
	{\includegraphics[width=0.22\textwidth]{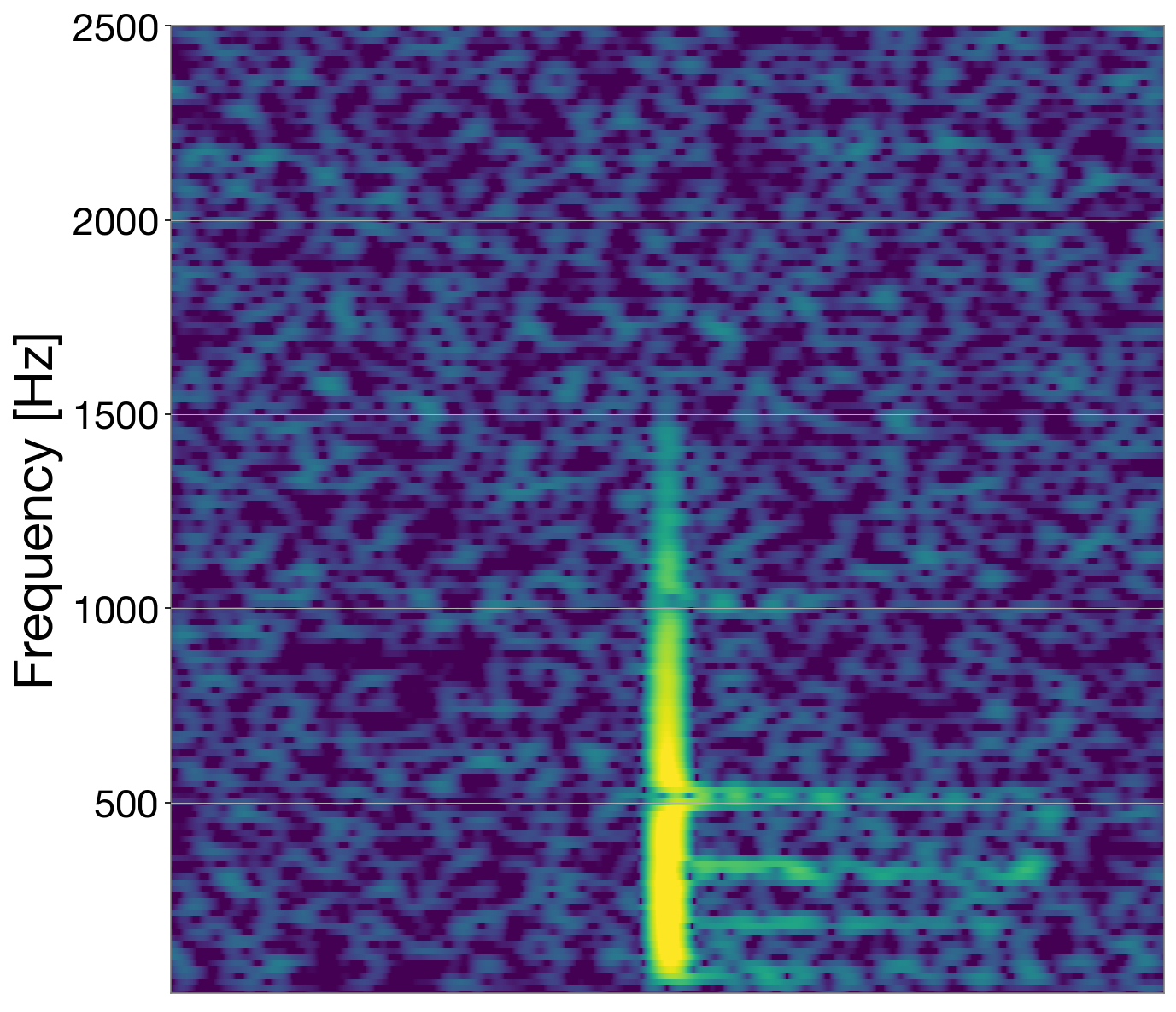}}\\
	
	{\includegraphics[width=0.22\textwidth]{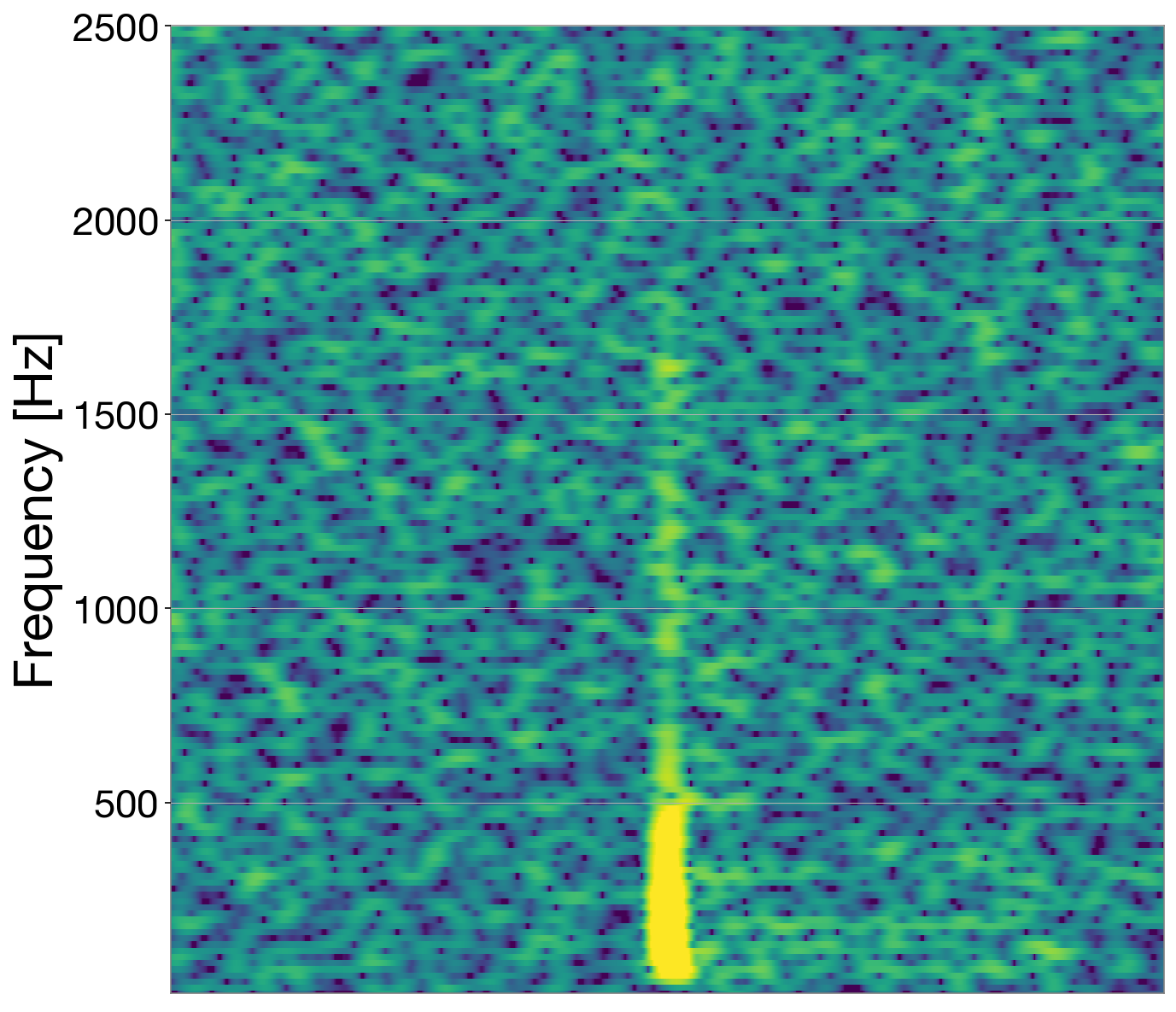}}
	{\includegraphics[width=0.22\textwidth]{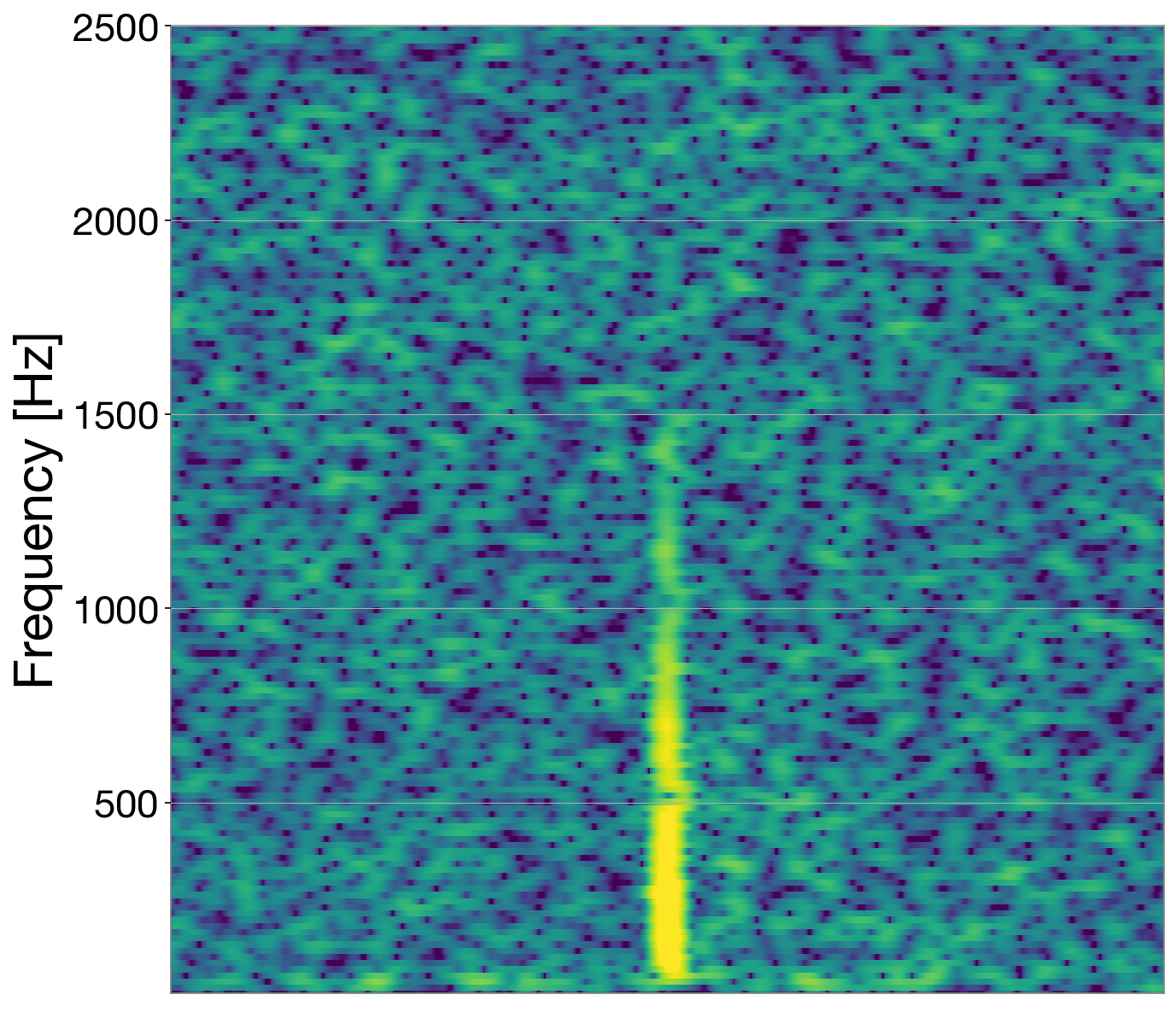}}
	{\includegraphics[width=0.22\textwidth]{./Figures/spect_single_7_original.png}}
	{\includegraphics[width=0.22\textwidth]{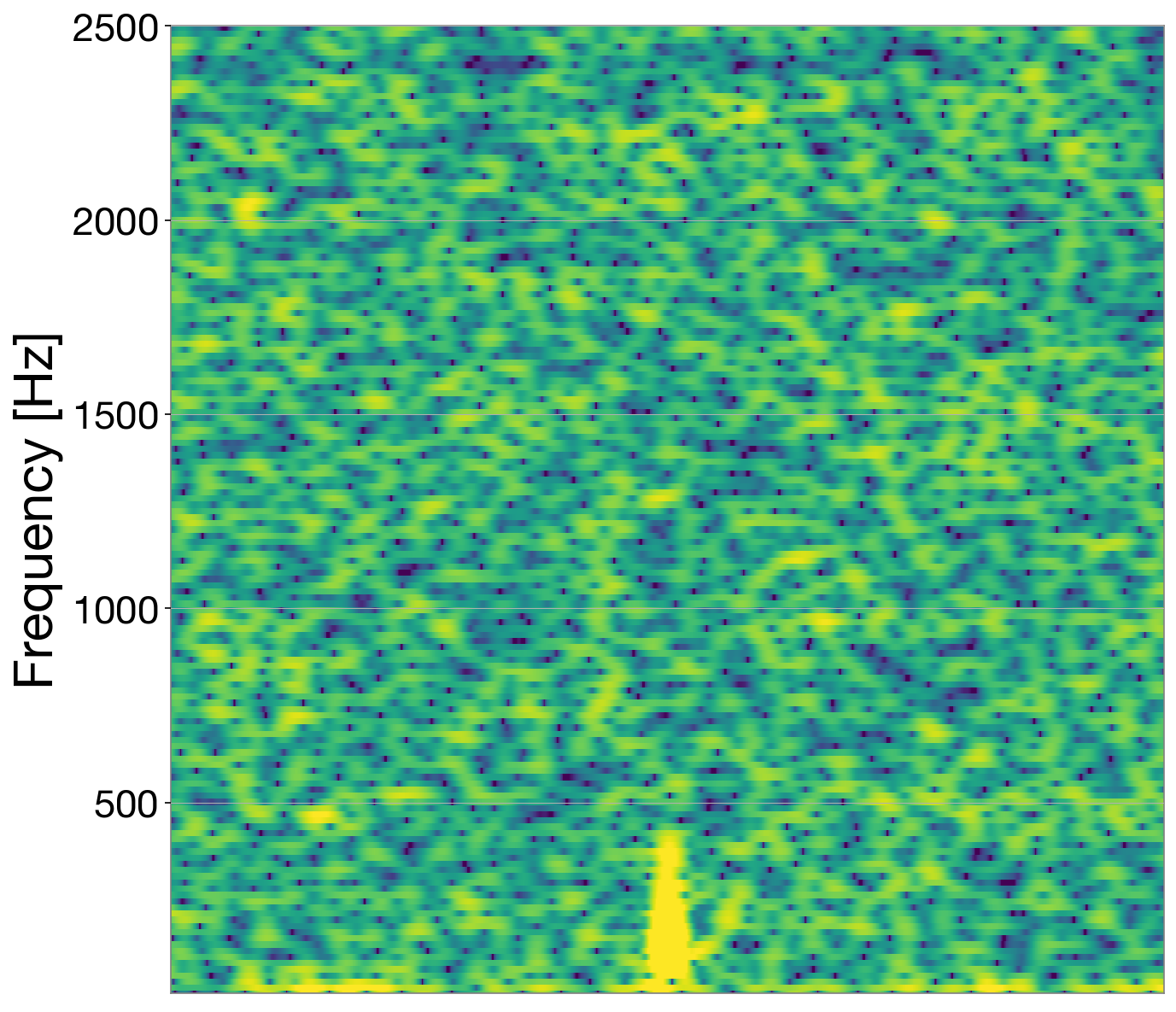}}\\
	
	{\includegraphics[width=0.22\textwidth]{./Figures/spect_single_9_original.png}}
	{\includegraphics[width=0.22\textwidth]{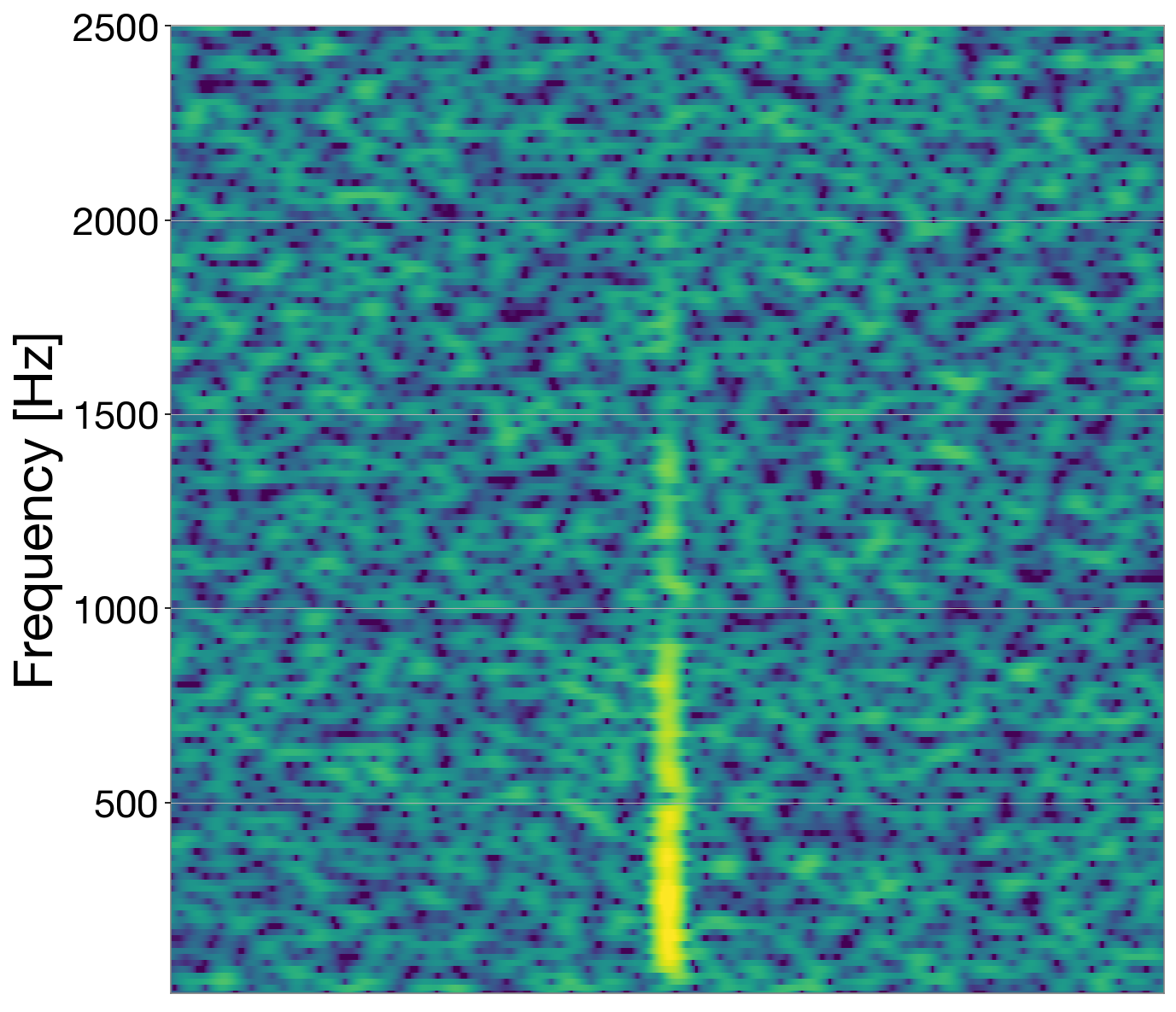}}
	{\includegraphics[width=0.22\textwidth]{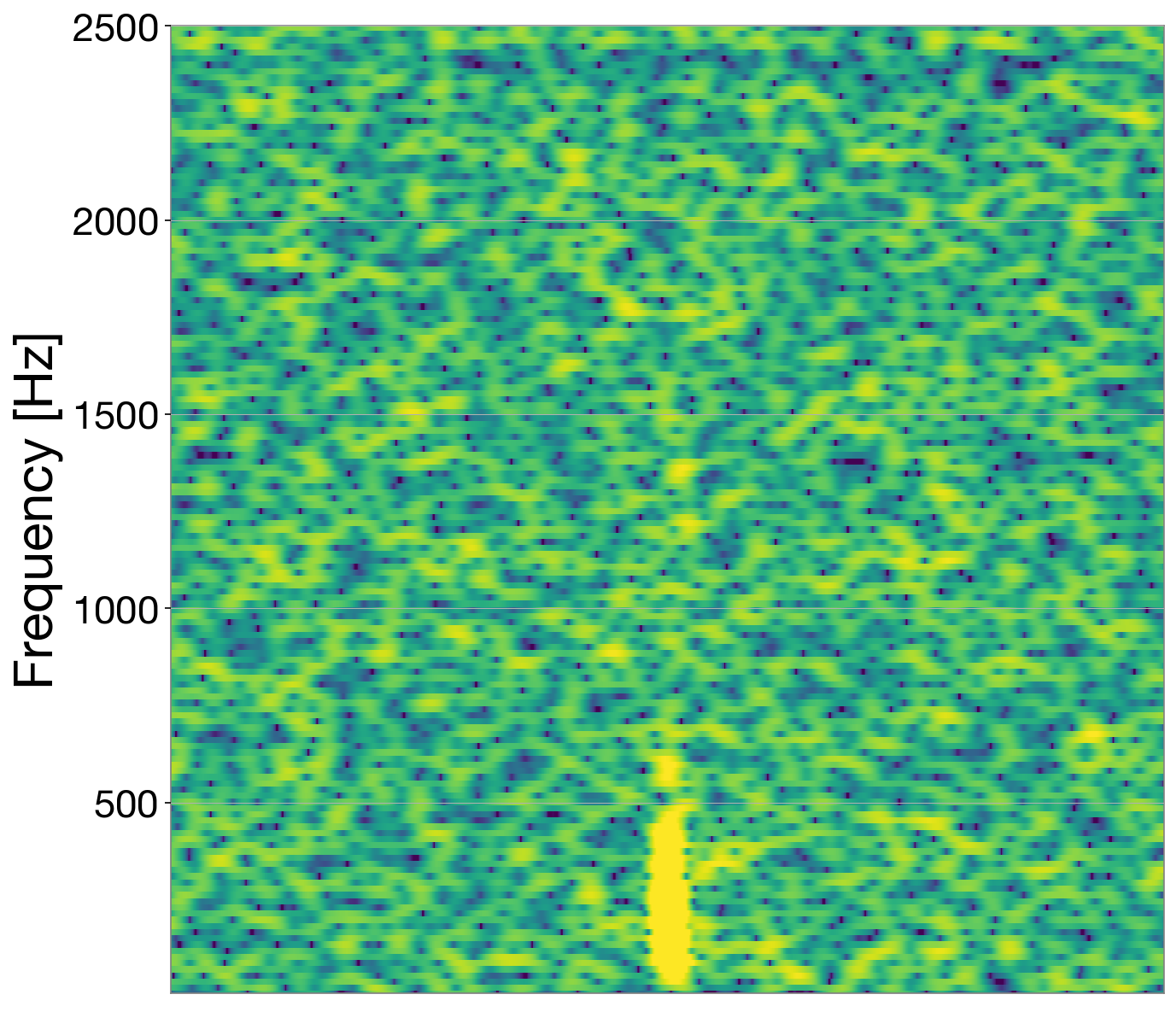}}
	{\includegraphics[width=0.22\textwidth]{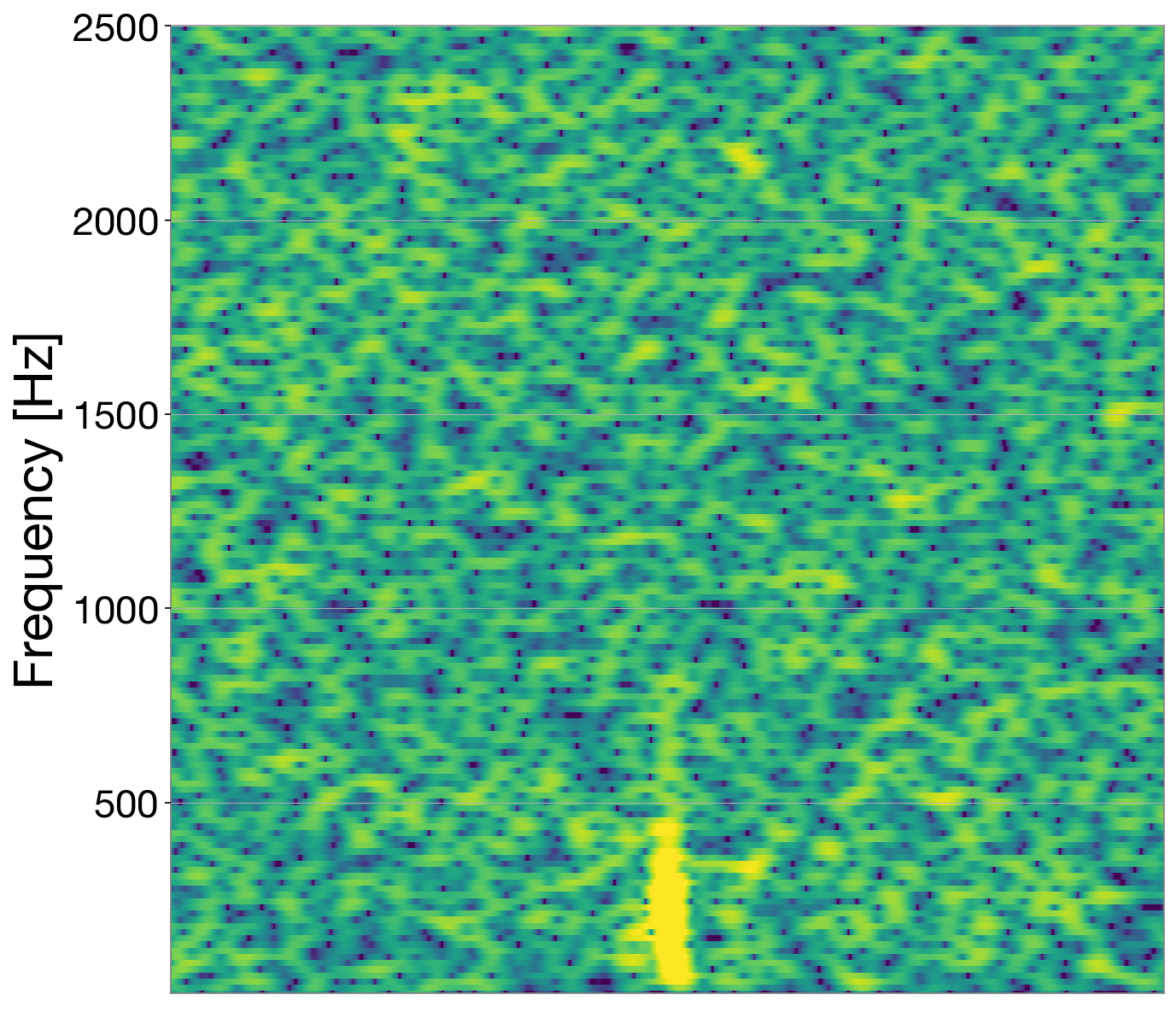}}\\
	
	{\includegraphics[width=0.22\textwidth]{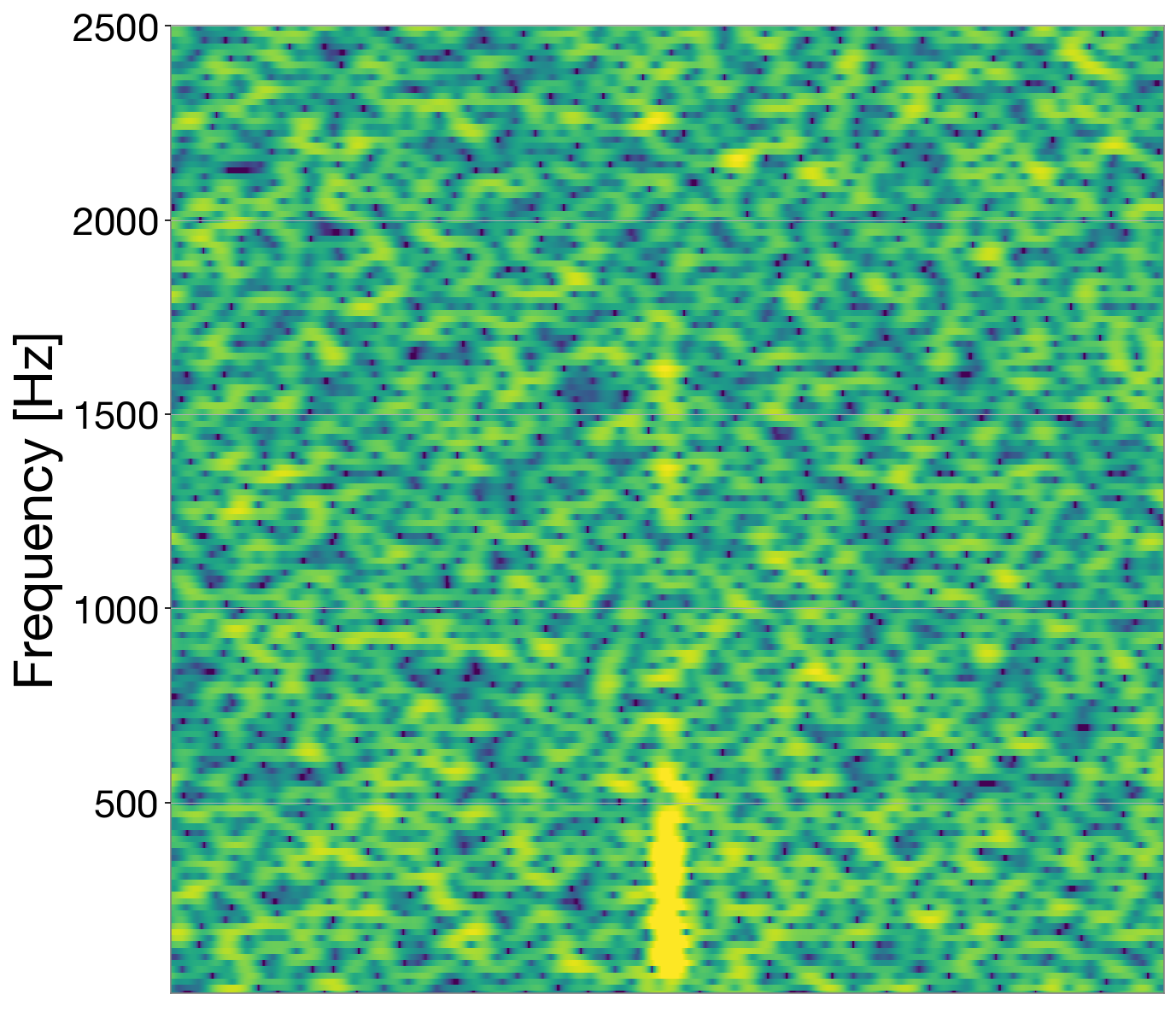}}
	{\includegraphics[width=0.22\textwidth]{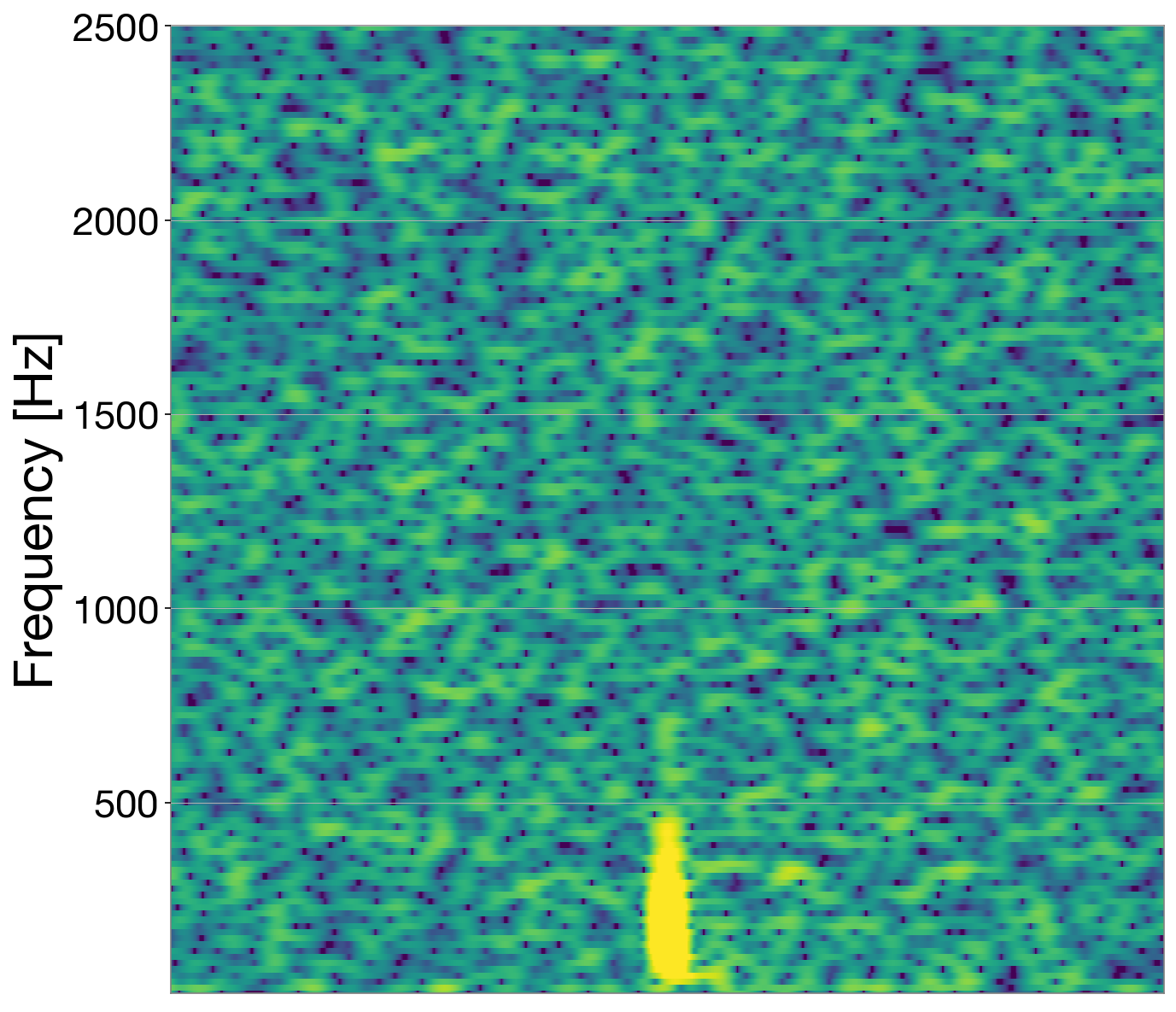}}
	{\includegraphics[width=0.22\textwidth]{./Figures/spect_single_15_original.png}}
	{\includegraphics[width=0.22\textwidth]{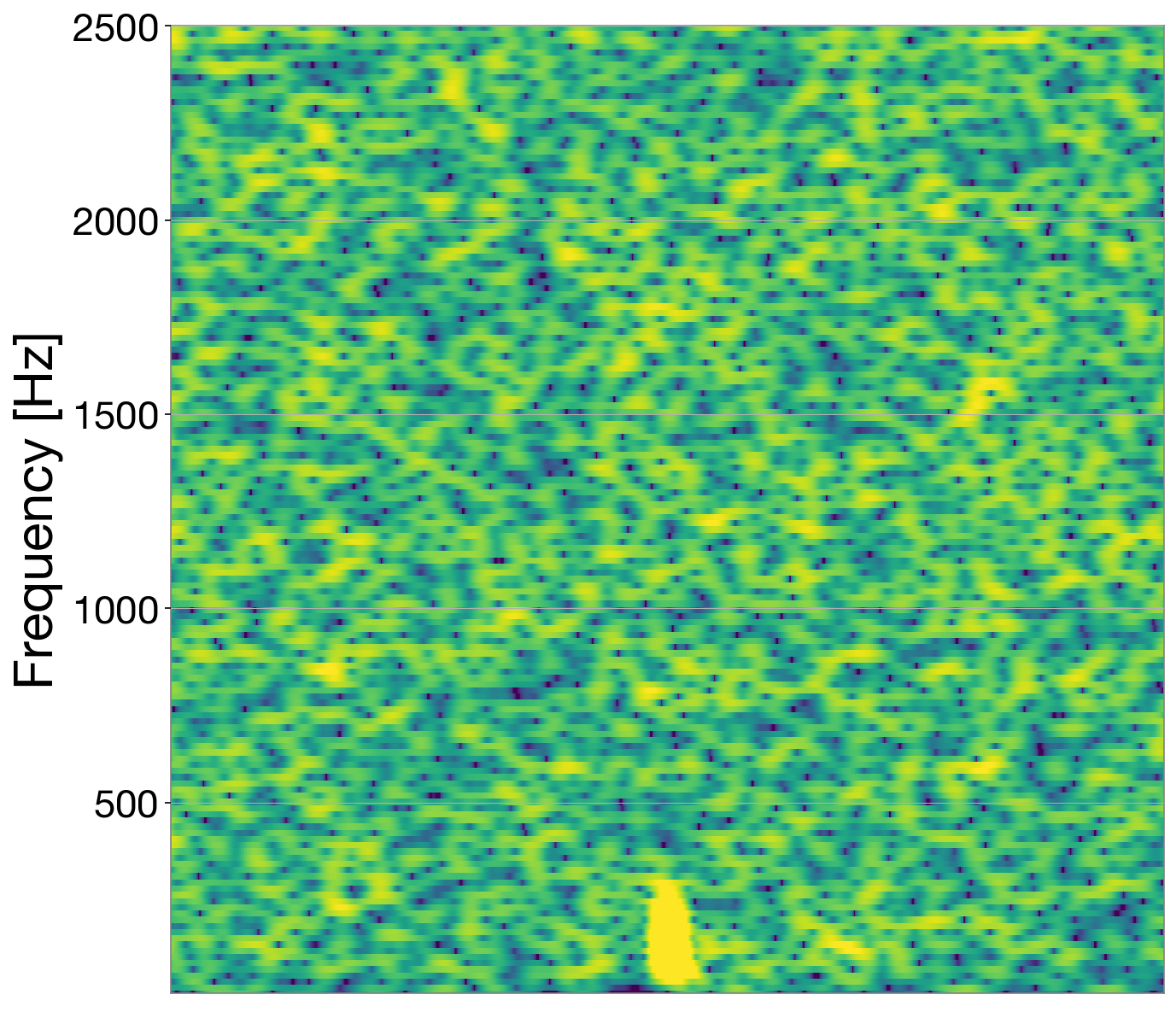}}\\
		
	 \caption{Time-frequency diagrams of all the blip glitches used as a test set.}
 \label{fig:all_blips_orig}
\end{figure*}

This appendix summarizes the main properties and time-frequency characteristics of the 16 blips used to test our algorithms. The physical parameters are reported in Table II while Fig.~\ref{fig:all_blips_orig} displays the time-frequency diagrams of all the 16 blips.

\begin{table*}
\label{tab:parameters}
\caption{Physical parameters of the 16 blip glitches of the test set. Note that the SNR values reported in the table are those provided by Gravity Spy~\cite{Zevin:2017} and are different to the values of our SNR estimator.}
\begin{center}
\begin{tabular}{ccccccccc}
Test \# & GPS time& Peak frequency& SNR&Amplitude& Central frequency &Duration&Bandwidth & IFO\\ 
& [s] & [Hz] & && [Hz] & [s] & [Hz] \\ 
\hline
1 & 1135873560.669 &   170.65 &   25.114 & 1.79e-22 &   294.23 &    0.375 &   521.87 & H1\\ 

 2 & 1126714444.138 &   339.88 &   33.524 & 6.89e-21 &   809.47 &    0.188 &   1554.9 & H1\\ 

 3 & 1127209450.153 &   149.41 &    35.98 & 5.12e-22 &   300.04 &    0.281 &   536.07 & L1\\ 

 4 & 1126792347.938 &    417.4 &   142.21 & 2.37e-19 &   1212.7 &    0.563 &   2361.5 & L1\\ 

 5 & 1135866057.468 &   211.48 &   58.114 & 4.34e-22 &   1926.7 &    0.625 &   3798.3 & H1\\ 

 6 & 1132400247.362 &   324.75 &   40.023 & 2.52e-21 &   1262.8 &    0.281 &   2457.2 & L1\\ 

 7 & 1135130271.231 &   211.48 &   18.767 & 1.77e-22 &   2421.1 &    0.248 &   4737.3 & L1\\ 

 8 & 1132810050.112 &   137.71 &   29.714 & 2.75e-22 &   298.75 &    0.375 &   512.83 & L1\\ 

 9 & 1127455791.581 &     1166 &   40.429 & 1.47e-20 &   1563.5 &    0.375 &   3063.1 & H1\\ 

10 & 1132779741.205 &   402.44 &   27.316 & 1.98e-20 &   2391.7 &    0.263 &   4698.8 & L1\\ 

11 & 1135129990.751 &   211.48 &    30.26 & 2.83e-22 &   720.44 &    0.375 &   1372.6 & L1\\ 

12 & 1135153845.843 &   262.06 &   26.046 & 5.26e-22 &   440.05 &    0.156 &   824.97 & L1\\ 

13 & 1131949939.534 &   324.75 &   19.041 & 5.65e-22 &   1934.1 &    0.094 &   3783.6 & H1\\ 

14 & 1126795399.374 &    183.5 &    50.77 & 7.09e-22 &   444.39 &    0.313 &   824.77 & L1\\ 

15 & 1127054242.325 &   149.41 &   53.247 & 4.29e-22 &   1232.6 &    0.313 &   2386.5 & H1\\ 

16 & 1135854885.481 &   137.71 &   30.468 & 2.57e-22 &   194.54 &    0.625 &   333.95 & H1\\ 

\hline
\end{tabular}
\end{center}
\label{tab:burst}
\end{table*}

\bibliographystyle{apsrev}
\bibliography{references}

\end{document}